\documentclass[%
 aip,
 amsmath,amssymb,
reprint,%
]{revtex4-1}

\usepackage{graphicx}
\usepackage{amsmath}
\usepackage{etoolbox}
\usepackage[version=3]{mhchem} 
\usepackage{wasysym}
\usepackage{amsmath,amssymb,latexsym}
\usepackage{pifont}
\usepackage[dvipsnames]{xcolor}

\begin{document}

\title[]
  {The First Direct Detection of Kirkwood Transitions in Concentrated Aqueous Electrolytes using Small Angle X-ray Scattering}
  
\author{Mohammadhasan  Dinpajooh}
\email{hadi.dinpajooh@pnnl.gov}
\affiliation
{Physical and Computational Sciences Directorate, Pacific Northwest National Laboratory, Richland WA}
\author{Elisa  Biasin}
\affiliation
{Physical and Computational Sciences Directorate, Pacific Northwest National Laboratory, Richland WA}
\author{Christopher J. Mundy}
\affiliation
{Physical and Computational Sciences Directorate, Pacific Northwest National Laboratory, Richland WA}
\affiliation
{Department of Chemical Engineering, University of Washington, Seattle WA}
\author{Gregory K. Schenter}
\affiliation
{Physical and Computational Sciences Directorate, Pacific Northwest National Laboratory, Richland WA}
\author{Emily T. Nienhuis}
\affiliation
{Physical and Computational Sciences Directorate, Pacific Northwest National Laboratory, Richland WA}
\author{Sebastian T. Mergelsberg}
\affiliation
{Physical and Computational Sciences Directorate, Pacific Northwest National Laboratory, Richland WA}
\author{Chris J. Benmore}
\affiliation
{Advanced Photon Source, Argonne National Laboratory, Chicago IL}
\author{John L. Fulton}
\email{john.fulton@pnnl.gov}
\affiliation
{Physical and Computational Sciences Directorate, Pacific Northwest National Laboratory, Richland WA}
\author{Shawn M. Kathmann}
\email{shawn.kathmann@pnnl.gov}
\affiliation
{Physical and Computational Sciences Directorate, Pacific Northwest National Laboratory, Richland WA}

\begin{abstract}
We provide a quantitative approach to describe electrolyte phenomena beyond Debye-H\"{u}ckel theory. Ion-ion correlations, screening, and equilibrium bulk structure
in various concentrated electrolytes are investigated using synchrotron
small angle X-ray scattering (SAXS), theory, and molecular simulation. Using  SAXS measurements we provide the first direct detection of the Kirkwood Transition (KT) for a variety of aqueous electrolytes (\cf{NaCl}, \cf{CaCl2}, \cf{SrCl2}, and \cf{ErCl3}). The KT may be defined as the concentration above which the ion-ion correlations
 cease to decay exponentially with a single length scale given by the Debye length $\lambda_{\rm D}$, and instead decay as a damped oscillator with two length scales: an inverse screening length $a_0$ reflecting the decay of the electric potential and
an oscillation frequency, $Q_0=2\pi/d$, that reflects the ordering in ion-ion correlations. The nonlocal electric susceptibility $\chi(Q)$ is also calculated which provides further insight into the spatial response to external fields. We find that the idealized definition of the KT needs to be generalized as our analyses show that the concentration dependencies of $a_0$ and $Q_0$ for our 1:1, 2:1, and 3:1 electrolytes are different.
\end{abstract}


\maketitle 

\section{Introduction}
Understanding screening and structure in concentrated
electrolytes has remained at the forefront of condensed phase statistical mechanics beginning nearly 100 years ago with the 1923 development of Debye-H\"{u}ckel theory (DH)\cite{DebyeHuckel1923}. This was after Debye's 1915 ``Scattering of X-rays" paper that played a fundamental role in the development of scattering from liquids\cite{Debye1915}. By 1927, X-ray scattering measurements by Prins and others\cite{ZernikePrins1927,Prins_1_1929,Prins_2_1929,Prins1934,Prins1935,PrinsFonteyne1935} observed prepeaks in the scattered intensity of aqueous electrolytes (\emph{e.g.,} \ce{Th(NO3)4} and \ce{UO2(NO3)2}) and concluded they arose from intermediate range order due to the ions and accompanying hydrating water molecules. In 1936 Kirkwood\cite{Kirkwood1936} predicted that oscillations would appear in the charge-charge spatial correlations as the concentration increased, the \emph{Kirkwood transition} (KT), in contrast to the DH exponential decay. Subsequent electric potential theories\cite{Singer1975,Kjellander1995,LeeFisher1997,Adar2019} to improve upon DH have been developed for finite-sized ions and yield similar KT behavior. Later, neutron scattering experiments by Enderby \emph{et al.}\cite{Howe1974,Neilson1975} found that the position of the prepeak for aqueous \ce{NiCl2} varied linearly $c^{1/3}$, where $c$ is the concentration, consistent with more recent studies\cite{Gaspar2004,Ribeiro2005,Fetisov2020,Ramamoorthy2020}. Using the Ornstein-Zernike (OZ) equation with the hypernetted chain approximation (HNC) for primitive and mean spherical ions, Attard\cite{Attard1993}, Leote De Carvalho and Evans\cite{Leote1994,Evans1994}, and Warren \emph{et al.}\cite{Warren2013} found non-Debye screening lengths \emph{i.e.,} the screening lengths increasing with concentration. A phenomenon referred to as \emph{underscreening}. Other Landau functional free energy theories have incorporated spatial and charge frustration into reduced models (\emph{e.g.} effective field theories and lattice models) predict the emergence of rich phase behavior\cite{TeubnerStrey1987,DeemChandler1994,LevinMundyDawson1992,Henderson1992,Bazant2011,Limmer2015,Rotenberg2018}. Krucker-Velasquez and Swan\cite{Krucker-Velasquez2021} using Brownian Dynamics with primitive ions found non-Debye screening and oscillation frequencies consistent with the earlier electric potential approaches.

Recently, surface force apparatus (SFA) experiments by Gebbie \emph{et al.}\cite{Gebbie2013}, Smith and Lee \emph{et al.}\cite{SmithLeePerkin2016,LeeSmithPerkin2017}, Ducker \emph{et al.}\cite{Gaddam2019}, Han \emph{et al.}\cite{Han2018}, and atomic force microscopy (AFM) measurements by Hjalmarsson \emph{et al.}\cite{Hjalmarsson2017}, and combined SFA and AFM studies by Baimpos \emph{et al.}\cite{Baimpos2014} found large \emph{effective} decay lengths mostly with ionic liquids but also with aqueous LiCl, NaCl and CsCl. However, their effective decay lengths were anomalously large compared to those found from liquid state theory, simulation, and SAXS experiments\cite{Cats2021,Zeman2020,Gavish2018,Adar2019,Rotenberg2018,Fetisov2020,Duignan2021}. In the SFA and AFM studies, they compared their decay lengths, by ansatz, to the Debye screening lengths. This discrepancy in measuring and quantifying screening lengths is due to a variety of factors. The mathematical description of the SFA force curve is based on a phenomenology that includes viscous, van der Waals, electrostatic, and solvation forces\cite{Israelachvili-1988,Evans-1999,SmithLeePerkin2016,LeeSmithPerkin2017} and exactly how these measurements are related rigorously to the statistical mechanics of the underlying ion-ion correlations have yet to be developed,\cite{Kardar-1992,Kardar-2015,Seyedi-2019} and a rigorous treatment will involve both nonlocal and frequency responses.\cite{Parsegian-1972,Ninham-1976,Parsegian-book,Klimchitskaya-2009} In fact, SAXS and SANS measurements on similar ionic liquids by Cabry \emph{et al.},\cite{Cabry-2017,Cabry-2022} using the same scattering distributions discussed in the present paper, do not find the anomalously large decay lengths found with SFA. To date, no other experiments or molecular simulations have confirmed anomalously large screening lengths.\cite{Coles-2020,Rotenberg2018,Rotenberg-2018} Furthermore, the most recent AFM measurements on a Li, Na, and Cs chloride salt solutions from 1 mM to 5 M found no anomalous increase in screening length.\cite{Mugele-2022} The screening length is defined as the asymptotic decay of an ion's electric potential due to the interactions with other ions and water molecules in the surrounding solution and measurements or calculations must be appropriately designed in order to be compared with the DH screening length.

Although both spatial and charge 
frustrations have
similar origins, namely the competition between short- and long-range interactions,
charge frustration produces unique and subtle signatures via charge correlations that
are not present when considering only mass density correlations. The theoretical underpinnings of non-Debye screening and oscillation frequency arise from the strong Coulombic interactions and subsequent correlations, a clear and quantitative relationship between model systems, theory, and experimental observations has remained elusive. Currently, it is impossible to directly measure just the charge-charge correlations using scattering experiments. The relevance of charge-charge correlations to solution dielectric properties is that they take into account the spatial distribution of ions as well as the interactions between them. For example, the electrolyte non-local electric susceptibility $\chi(Q)$, which depends directly upon charge-charge correlations\cite{Kjellander2019,HansenMcDonald2013}, describes the response of a system to external fields. Understanding and controlling the ions valences, solution concentration, and temperature allow the use of external fields to induce changes inside the electrolyte and thereby allow active control of the systems response (\emph{e.g.,} electrodeless electrolysis, liquid antennas, catalysis, crystallization pathways and rate, colloid interactions, transport properties, etc.)\cite{Mahdisoltani2021}.

The response charge density due to ion $j$ 
\begin{equation}
    \rho_j(r)=\sum_i n_i z_i e g_{ij} (r) 
    \label{eq:charge}
\end{equation}
where  $i,j\rightarrow +,-$ is defined through the corresponding radial distribution functions $g_{ij}(r)$ between electrolyte ions. The KT separates the isotropic liquid state into two 
distinct idealized regions shown in Figure~\ref{fig:KT_kjell}: (\emph{1}) Where the charge correlations decay exponentially (\emph{EXP}) with a single length scale denoted $a_0^{-1}$ herein via the total correlation function $h_{ij}(r)=\frac{A}{r}\exp(-a_0 r)$, where $h_{ij}(r) = g_{ij}(r)-1$ and (\emph{2}) 
a region where two length-scales are dominant (denoted $a_0$ and $Q_0$ herein via a damped oscillator (\emph{DOSC}) 
$h_{ij}(r)=\frac{B}{r}\exp(-a_0 r)\cos(Q_0 r - \delta)$), where $Q_0$ is referred to as the periodicity or spatial oscillation frequency, and $\delta$ is a spatial phase. Stated differently, the conceptual possibility of a KT could be arrived at starting from DH theory simply by allowing the inverse screening length $\kappa$ in the exponential decay $\frac{A}{r}e^{-\kappa r}$ to become complex \emph{i.e.}, $\kappa=a_0+iQ_0$, and thus $\frac{A}{r}e^{-\kappa r}\rightarrow \frac{B}{r}e^{-a_0r}cos[Q_0r-\delta]$. 
In the exponential regime, $a_0\rightarrow\kappa_{D}$, where $\kappa^2_{\rm D}=\frac{4\pi e^2}{\varepsilon k_{B}T}\sum_i n_i z_i^2$, and the Debye length $\lambda_{\rm D}=1/\kappa_{\rm D}$.
Here $e$ is the electron charge, $n_i$ is the number density of the $i^{\rm th}$ ion species, $\varepsilon$ is the dielectric constant of the solvent, and $z_i$ is the valence charge of the $i^{th}$ ion species. Thus, at the KT two length scales arise: a screening length $\lambda=1/a_0$ distinct from the Debye length $\lambda_{\rm D}$, and an oscillation frequency $Q_0=2\pi/d$ where $d$ is the correlation length between ions.

The physical picture here is that at infinite dilution where DH theory is valid, the ions simply do not see one another and there is only a single screening length over which each ion's electric potential exponentially decays into the surrounding water. As the number of ions is increased (and DH theory breaks down), the ions begin to interact with each other through the intervening water and their screening length is reduced. At the KT, the electric potential decays as a damped oscillator where each ion is surrounded by an alternating atmosphere with decreasing amplitudes of positive and negative charges and the screening length increases. In this way, as a dilute electrolyte becomes more concentrated, the screening length drops from infinity to a minimum and then starts to gradually increase again while the alternating ion atmosphere extends further out into the electrolyte. 

In Small Angle X-ray Scattering (SAXS) the total scattering is due to the sum of all pair correlations (\emph{e.g.} including those with water) weighted by their atomic X-ray scattering factors. As a result, the individual charge-charge correlations, that are of most interest here, interfere with other pair correlations. Using the 
theoretical connection with the SAXS signal one can then access the charge correlations to make \emph{quantitative} comparisons between screening lengths and periodicity as a 
function of ionic strength and electrolyte type.  Indeed, this approach was 
undertaken in a recent study where we quantified the origin of the observed 
prepreaks in 2:1 electrolytes as a function of concentration\cite{Fetisov2020}. This study demonstrated that it is possible to obtain \emph{quantitative} agreement with SAXS using a class of Kirkwood-Buff derived Force Fields\cite{Naleem2018} (KBFFs) that are 
fit to reproduce the experimentally observed concentration dependence of isothermal compressibilities, partial molar volumes, and activities. The physical picture that emerged from this study is that the prepeaks in the SAXS can encode the charge-charge correlations, and hence $a_0$ and $Q_0$, for various electrolyte solutions.

As mentioned previously, recent measurements of non-Debye screening lengths in concentrated electrolytes was investigated by Perkin \emph{et al.}\cite{LeeSmithPerkin2017,SmithLeePerkin2016} using a Surface Force Apparatus (SFA), Ducker \emph{et al.}\cite{Gaddam2019} using thin film interference/fluorescence microscopy (IFM), and Fetisov \emph{et al.}\cite{Fetisov2020} using SAXS. In all these measurements, in contrast to the DH screening behaviour $\lambda_{\rm D}(c)$, the screening length $\lambda(c)$ was found to increase with concentration typically starting at about $c\approx0.5$ m. This non-Debye behaviour is referred to as underscreening in the sense that the spatial extent of an ion's influence out into the surrounding solution, as quantified by the electric potential response, is larger than $\lambda_{\rm D}$. However, the SFA and IFM $\lambda(c)$ are anomalously large at the higher concentrations where, for example, $\lambda_{\rm SFA}\approx33$\,\AA\,and $\lambda_{\rm IFM}\approx65$\,\AA\, were found for \ce{NaCl} at 5 M (compared to $\lambda_{\rm D}\approx1.4$\,\AA. Our own mean field \ce{NaCl} OZ/HNC calculation $\lambda_{\rm HNC}\approx2-3$\,\AA). These experimental estimates of ion-ion correlation decay lengths assume they probe only the bulk electrolyte and hence are insensitive to the surfaces employed. The SAXS measurements used in the present investigation provide a more direct probe and statistical mechanical connection with the underlying correlations in electrolytes than provided by SFA or IFM measurements. Furthermore, SAXS yields information on \emph{both} the screening length and oscillation frequency of the correlations. In our previous SAXS measurements on bulk electrolytes, supported by extensive molecular dynamics simulations, found $\lambda_{\rm SAXS}\approx5.7$\,\AA\ for \ce{CaCl_{2}} at 4 m (compared to $\lambda_{\rm D}\approx0.7$\,\AA). 

\begin{figure}[tbh]
\centering
\includegraphics[width=\columnwidth]{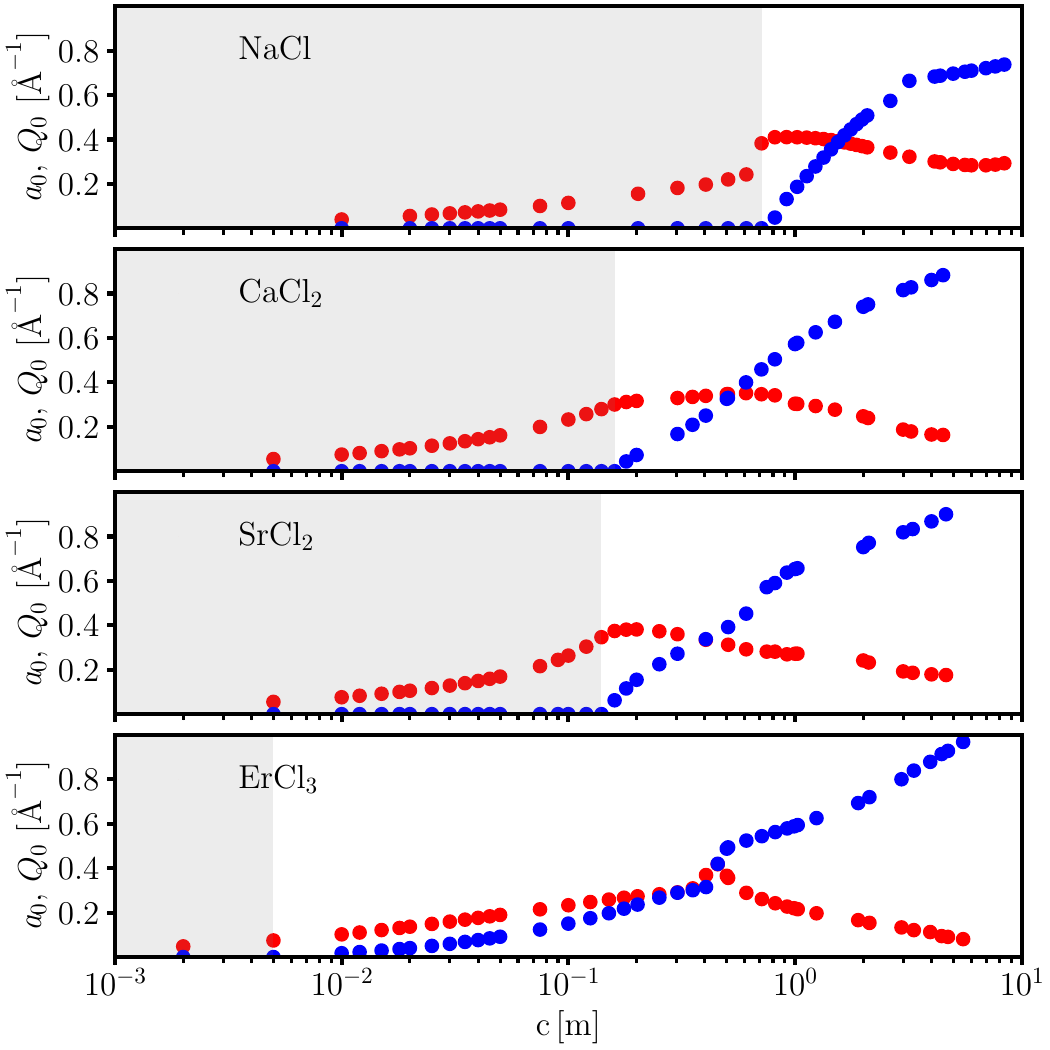}
\caption{KT regions for various electrolytes with increasing cation valence. The inverse screening lengths $a_0=1/\lambda$ (red circles) and periodicities $Q_0=2\pi/d$ (blue circles) as described by HNC with mean field ion-ion interactions. The shaded area shows where $Q_0 \rightarrow 0$.}
\label{fig:All-a0Q0}
\end{figure}

Herein, we connect the liquid state theory and molecular simulations to experimental SAXS (a.k.a. \emph{``The Grand Dilution"}) for a variety of aqueous electrolytes with atomic cations of varying valence (\emph{e.g.}, +1, +2, and +3). Because of the long-length scale nature of non-Debye screening lengths and periodicity phenomena, reduced descriptions \emph{e.g.} charged hard-spheres in a continuum dielectric and lattice models are able the capture the qualitative features of screening and periodicity as a function of concentration. 
A remaining challenge is to connect all-atom molecular simulations to the reduced representations. 
In this work, theoretical models for electrolyte solutions have been developed based on the liquid state theories using the ion-ion mean-field interactions in water, which involve accurate short- and long-range interactions and can produce quantitative agreement with all-atom molecular dynamics simulations. 
Figure~\ref{fig:All-a0Q0} presents the the predictions of screening and periodicity (root structure) with increasing concentration and valency of electrolyte solutions including (\ce{NaCl}, \ce{CaCl2}, \ce{SrCl2}, and \ce{ErCl3}) in which the inverse screening lengths and oscillation frequencies are extracted from the cation-cation asymptotic correlations. Such analyses show that the traditional KT pole structure holds reasonably well for \ce{NaCl} electrolyte solutions but as the valency of electrolyte solutions increases, a more complex root structure emerges and will be discussed further below. These results are supported by all-atom molecular dynamics simulations and are quantitatively consistent with our SAXS spectra.  

\section{Results and discussion}

\begin{figure}[tbh]
\centering
\includegraphics[width=\columnwidth]{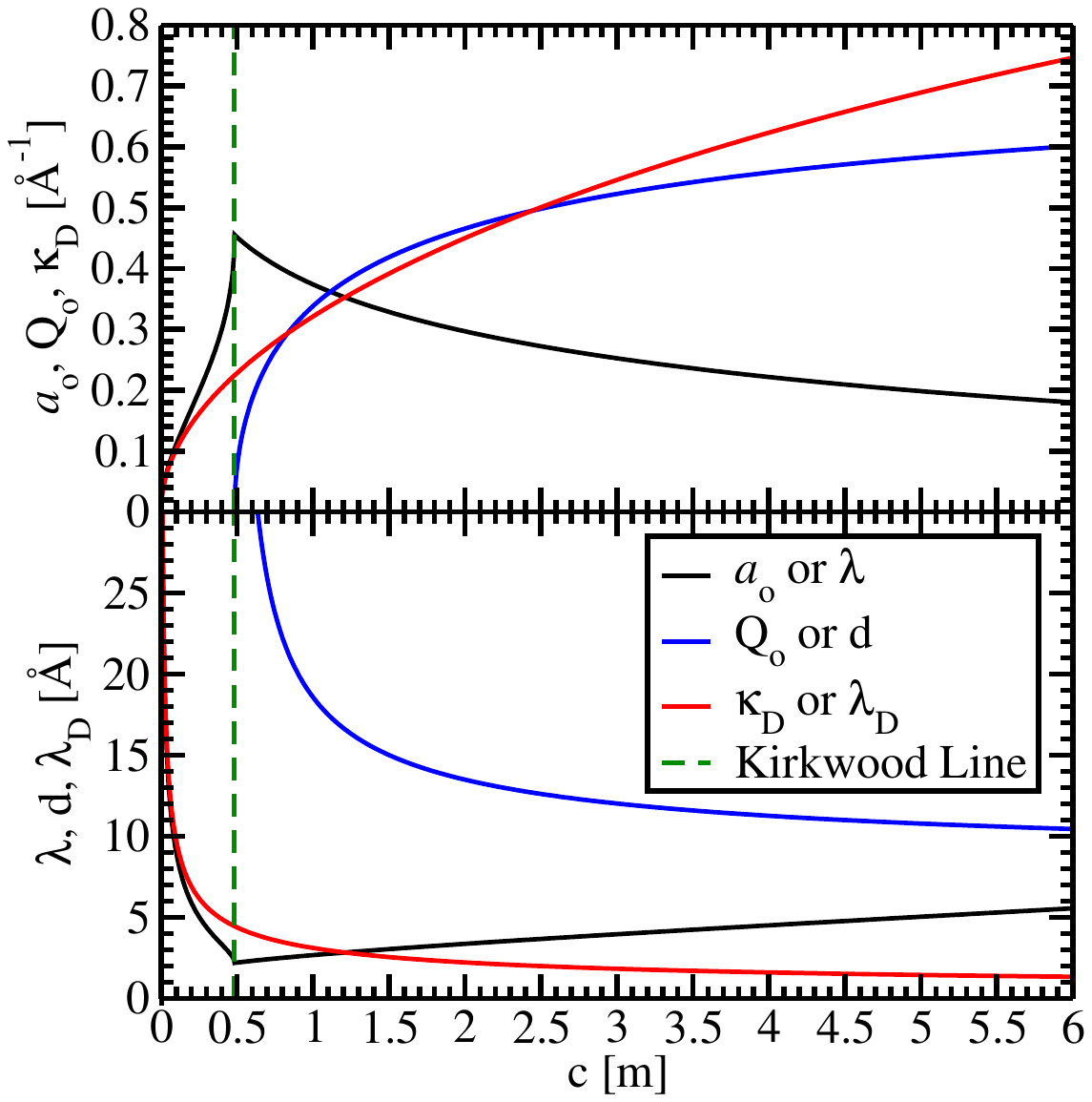}
\caption{Concentration dependent $a_0$ or $\lambda$ (black), $Q_0$ or $d$ (blue), and $\kappa_{\rm D}$ or $\lambda_{\rm D}$ (red), and  for aqueous \ce{NaCl} from numerical solution of $\kappa^{2}=\kappa_{\rm D}^{2}e^{z}/(1+z)$ with a radius of $6.0$\,\AA. Differently than what is predicted by Debye theory, a cusp occurs in the concentration dependence of $a_0(c)$, and $Q_0(c)$ rises at the KT (green dashed line).}
\label{fig:KT_kjell}
\end{figure}

As noted in the introduction, past theoretical approximations of the electric potentials for finite-sized ions yielded a variety of transcendental equations (TEs) that can be solved to obtain real and imaginary components: $a_0$ and $Q_0$, respectively. One of the simplest TEs, derived by Kjellander\cite{Kjellander1995} and based on an extension of the linearized Poisson-Boltzmann equation for a 1:1 electrolyte, is given by
\begin{equation}
    \kappa^{2}=\kappa_{\rm D}^{2}e^{z}/(1+z)
\end{equation}
where $\kappa=a_0+iQ_0$,\,$z=\kappa a$, and $a$ is the ion diameter. We solve this numerically using the Newton-Raphson method for aqueous \ce{NaCl} and the results are shown in Figure~\ref{fig:KT_kjell}. Kjellander's TE, as well as several other TEs from liquid state theory (all display a similar root structure and are provided in the SI for comparison), contains several important features to aid in understanding the screening and periodicity (or oscillation frequency) behavior of aqueous electrolytes arising from the underlying ion-ion correlations as well as their departure from DH behavior. One can see in Figure~\ref{fig:KT_kjell} that the idealized KT is characterized by a Kirkwood line (KL), indicated by the vertical dashed line (green), and is predicted to be $c\approx0.49\,m$, where a cusp occurs in the concentration dependence of $a_0(c)$, and $Q_0(c)$ rises from zero at the KT. For reference, $\kappa_{\rm D}$ is included to show that a deviation exists in the screening behavior well before the KT. 
Figure~\ref{fig:KT_kjell} also shows the screening lengths $\lambda(c)$, $\lambda_{\rm D}(c)$, and $d(c)$-spacing. It should also be noted that similar physical length scales appear $\lambda\sim6$\,\AA\,and $d\sim10$\,\AA\, as the electrolyte approaches saturation $c\approx6\,m$ as shown in Figure~\ref{fig:KT_kjell}.

We will see from the results presented below, that this idealized behavior of $a_0$ and $Q_0$ at the KL, or, said differently, how we \emph{define} and characterize the KT, \emph{e.g.}, with a Kirkwood line, may need to be generalized to a \emph{Kirkwood region} depending on the specific electrolyte.  

Let's consider how the inverse screening length $a_0$ and oscillation frequency $Q_0$ of ion-ion correlations can be probed using X-ray scattering. The Fourier transform (FT) of the total correlation function $h_{ij}(r)$ is
\begin{equation}
\hat{h}_{ij}(Q)=\frac{4\pi}{Q}\int_{0}^{\infty}rh_{ij}(r)\sin[Qr]dr,
\end{equation}
where $Q$ is the scattering wavevector with the structure factor given by 
\begin{equation}
S_{ij}(Q)=c_{i}\delta_{ij}+nc_{i}c_{j}\hat{h}_{ij}(Q).
\end{equation}
Taking the FT of the \emph{DOSC} for $h_{ij}(r)$
\begin{equation}
    h_{ij}(r)=\frac{Be^{-a_0r}}{r}\cos[Q_0r-\delta],
\end{equation}
one obtains the
Teubner-Strey (TS) distribution:
\begin{equation}
\begin{split}
&\hat{h}_{ij}^{\rm TS}(Q) = \\ 
&\frac{4\pi B\left[\left(a_0^{2}-Q_0^{2}+Q^{2}\right)\cos[\delta]+2a_0 Q_0\sin[\delta]\right]}{\left[a_0^{2}+Q_0^{2}\right]^{2}+2\left[a_0^{2}-Q_0^{2}\right]Q^{2}+Q^{4}},
\end{split}
\end{equation}
that displays a peak at $Q\neq0$, with the peak position given by
\begin{equation}
Q_{\rm peak}=\sqrt{Q_0^{2}-a_0^{2}-2a_0Q_0(\sec[\delta]+\tan[\delta])}.
\end{equation}
A Lorentzian (L) distribution peaked at $Q=0$ is given by the limit:
\begin{equation}
\lim_{Q_0,\delta\rightarrow0}\hat{h}_{ij}^{\rm TS}(Q)=\hat{h}_{ij}^{\rm L}(Q)=\frac{4\pi B}{a_0^{2}+Q^{2}},
\end{equation}
as one may have anticipated since the FT of the \emph{EXP}
form for $h_{ij}(r)=\frac{Be^{-a_0r}}{r}$ yields $\hat{h}_{ij}^{\rm L}(Q)$.

The measured X-ray intensity $I(Q)$ is related to the structure factor $S_{X}(Q)$
by
\begin{equation}
I(Q)\approx S_{X}(Q)-1=n\sum_{i,j}c_{i}c_{j}f_{ij}(Q)\hat{h}_{ij}(Q),
\end{equation}
where
\begin{equation}
f_{ij}(Q)=\frac{f_{i}(Q)f_{j}(Q)}{\left[{\displaystyle \sum_{i=1}^{N}}c_{i}f_{i}(Q)\right]^{2}},
\end{equation}
where $n$ is the total system number density, $c_{i}$ is
the atomic fraction of species $i$, $f_{i}(Q)$ is the X-ray form
factor for species $i$, and $N$ is the total number of atomic species in the system. X-rays scatter off an element's electron density whose scattering power is given by the form factor $f_{i}(Q)$ defined by the Fourier transform of the electron density. Thus, elements with higher electron densities have a greater scattering power that scale as the atomic number $Z$. SAXS spectra resolve spatial correlations on the order of $2\pi/Q$, which in our studies $Q$ ranges from $\sim3$ to $0.01$\,\AA$^{-1}$, thus allowing us to probe spatial correlations ranging from $\sim2$ to $628$\,\AA. The peak position of the TS distribution gives the spatial correlation length $d=2\pi/Q$, and the peak full width at half maximum (FWHM) $\Delta Q$ reflects the range of spatial order or spatial coherence length $\xi=2\pi/\Delta Q$.

To characterize the prepeak region of the SAXS spectra for a given
aqueous electrolyte, we must consider both the strength of the ion-ion
correlations as well as the experimental signal contrast of these
correlations against the ubiquitous water background. Given that we are interested in understanding the low-$Q$ prepeak region as a function of concentration, we must choose our aqueous electrolytes judiciously. For example, in our dilution experiments aqueous sodium chloride \ce{NaCl} is much more difficult to characterize the prepeak region than aqueous erbium chloride \ce{ErCl_{3}}. The reasons for this are twofold: (1) the scattering power of \ce{Na^{+}}
(10\emph{e}) and \ce{Cl^{-}} (18\emph{e}) ions are not too dissimilar from \ce{O} (8\emph{e}) atom in water, and (2) both \ce{Na+} and \ce{Cl-} ions are monovalent and thus do not yield as strong ion-ion correlations compared to the trivalent \ce{Er^{+3}} (65\emph{e}) ions. In our
previous SAXS and MD studies on aqueous \ce{CaCl_{2}} we found that the prepeak region was dominated by the \ce{Ca^{+2}}-\ce{Ca^{+2}} correlations consistent with the above discussion. Thus, eventhough there are ten partials, $N(N+1)/2=10$ for $N=4$ (e.g., for aqueous \ce{ErCl_{3}},
we have \ce{O}, \ce{H}, \ce{Er}, and \ce{Cl}), underlying the X-ray scattering signal, the major contribution comes from the strong cation-cation total correlation function thus justifying the use of a single TS distribution to characterize the measured SAXS prepeak which we denote simply as $\hat{h}^{\rm TSF}(Q)$ where the $ij$ subscripts have been dropped. We note also that as ions are added to a dilute electrolyte the structure of water will also be perturbed and the \ce{O}-\ce{O} correlation function will develop a prepeak, albeit weaker, but consistent with the cation-cation prepeak evolution.

An analysis of the prepeak structure using a broad range of concentrations can be used to obtain $a_0(c)$ and $Q_0(c)$ and then compared to the results of molecular dynamics, integral equation theory, and DH theory. Importantly, this opens the tantalizing possibility, explored presently, that the KT can be characterized by observing a transition in low-\textit{Q} prepeak SAXS spectra from a TS to a L distribution in a series of dilution experiments on various electrolytes. For each concentration, we find it useful to add a linear background to the TS distribution yielding a six-parameter $(B,a_0,Q_0,\delta,m,b)$ fit function given by $\hat{h}_{ij}^{\rm TSF}(Q)=\hat{h}_{ij}^{\rm TS}(Q)+mQ+b$. In this way, the SAXS signal provides a direct connection to the underlying ion-ion correlations, pole structure, as well as quantities derived from those correlations.

Our previous investigations on reduced representations of ion-ion interactions focused on the precise nature of short-range interactions to accurately predict collective phenomena of concentrated electrolytes.\cite{Duignan2021} In these studies, the short-range (SR) interaction is computed using quantum density functional theory (qDFT) and the long-range (LR) interaction is modeled as $q_iq_j/\varepsilon r$. These studies suggest that simple charge-frustrated interactions (\emph{e.g.} point charge models in conjunction with a repulsive core) do not have the flexibility to describe SR phenomena as probed by estimates of the single ion solvation free energy and accurate X-ray absorption fine structure and X-ray absorption near-edge spectroscopy. The importance of accurate SR interactions has profound impact on our understanding of non-ideal properties of electrolytes such as clustering and pathways to nucleation of the solid polymorph.\cite{Henzler-2018} The additional experimental observation of prepeaks in SAXS provide an additional constraint on the important coupling of SR interactions to LR collective phenomena. Moreover, the presence of a region within the isotropic liquid phase with two dominant length-scales, at concentrations above the KL,  provide the necessary fluctuations that yield complex pathways to amorphous phases or crystal polymorphs that may be distinct from the well-characterized final crystalline phase.

\subsubsection{SAXS, MD, and OZ/HNC}
\begin{figure}[tbh]
\centering
\includegraphics[width=\columnwidth]{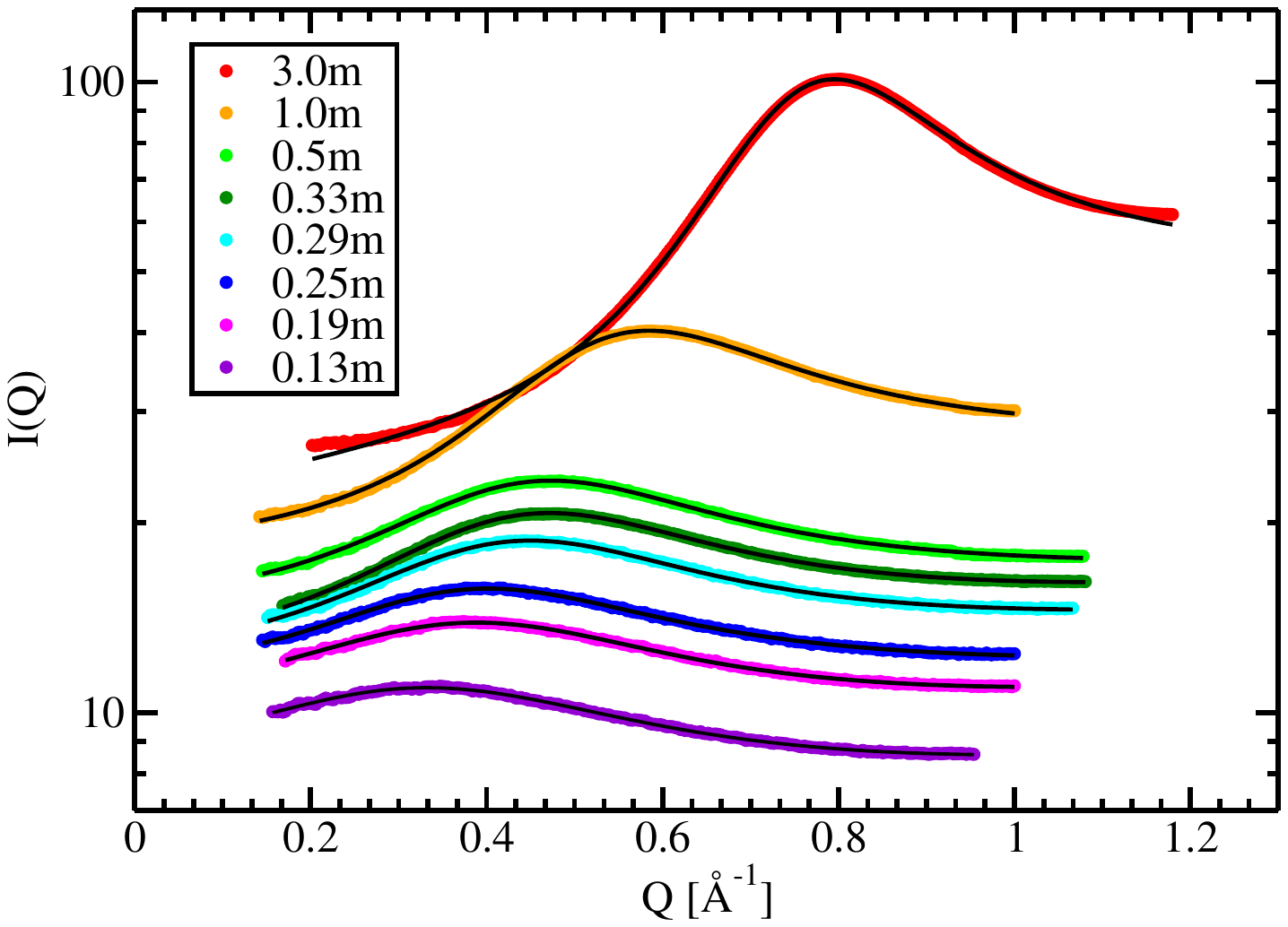}
\caption{A comparison of the experimental SAXS spectra (colors) and fits (black) for dilution series for aqueous \ce{ErCl_{3}} using the TS fitting distribution $h^{\rm TSF}(Q)$ to obtain $a_0=1/\lambda$ and $Q_0=2\pi/d$.}
\label{fig:ErCl3_expt_fit}
\end{figure}

Figure~\ref{fig:ErCl3_expt_fit} shows representative experimental SAXS data for the prepeaks and their fits using $\hat{h}^{\rm TSF}(Q)$ for aqueous \ce{ErCl3} for a range of concentrations. The fidelity of using a \emph{single} TS distribution is validated by how accurately it reproduces the experimental spectrum. We'll see later that a single dominant pole describing the ion-ion correlations can greatly simplify the development of reduced models. From these SAXS spectra, we obtain the following coherence lengths $\xi=2\pi/\Delta Q$, where $\Delta Q$ is the Full Width Half Maximum (FWHM) of the prepeak (these are quantitatively consistent with sizes deduced from the Scherrer equation), at the concentrations in Figure~\ref{fig:ErCl3_expt_fit}: 27.4\,\AA\,(3 m), 23.6\,\AA\,(1 m), 18.0\,\AA\,(0.5 m), 18.8\,\AA\,(0.33 m), 18.0\,\AA\,(0.29 m), 16.4\,\AA\,(0.25 m), 15.6\,\AA\,(0.19 m), 14.8\,\AA\,(0.13 m). These nanometer-sized coherence lengths increasing with concentration are consistent with the extent of the ion atmospheres discussed later.

\begin{figure}[tbh]
\centering
\includegraphics[width=\columnwidth]{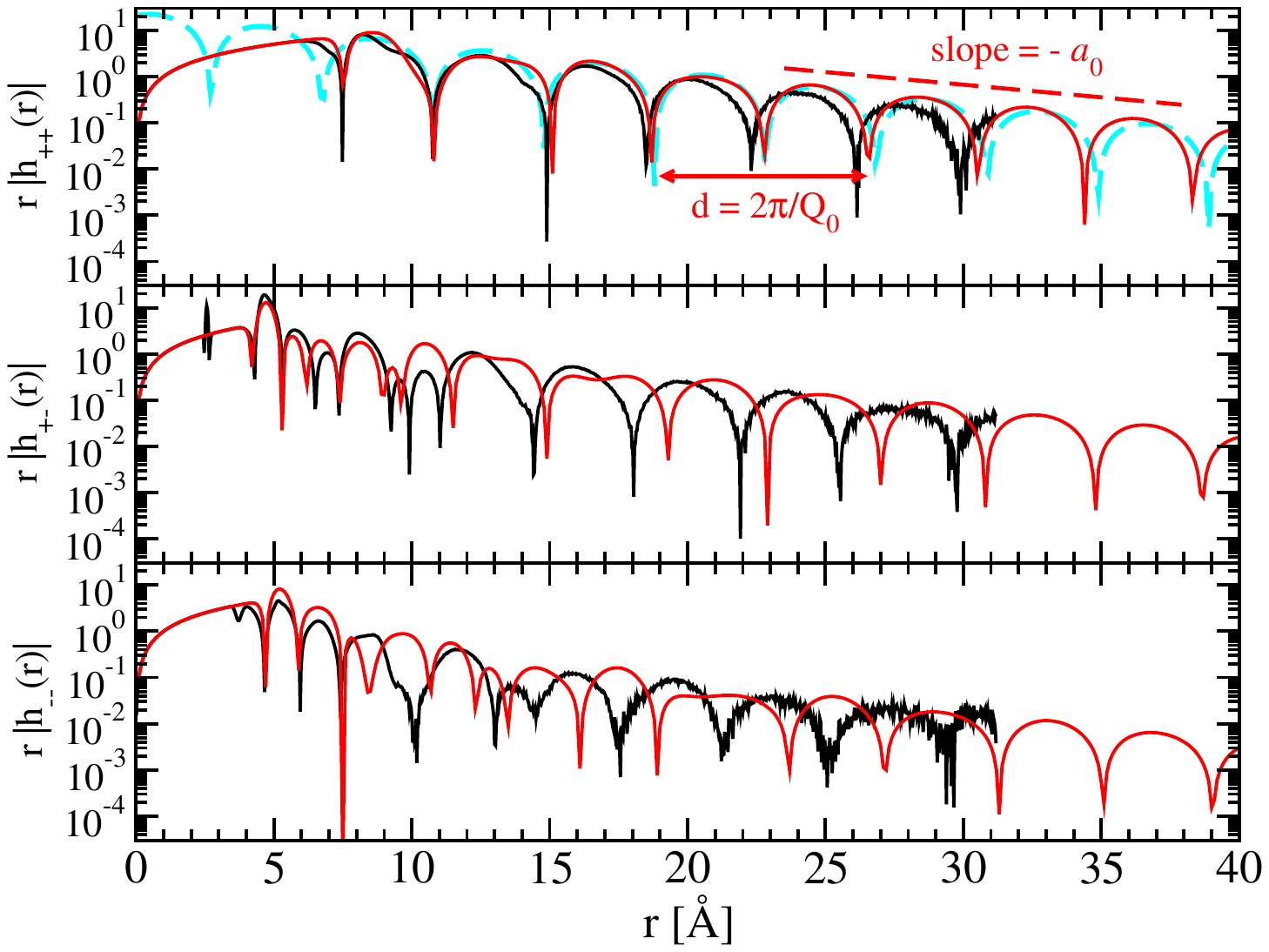}
\caption{Comparison of $r|h(r)|$ (in units of \AA) for Er-Er (top panel), Er-Cl (middle panel), and Cl-Cl (bottom panel) correlations from MD (black) and HNC (red) at $c=3$ m. For reference, the slope = $-a_0$ (red dashed line) and $d=2\pi/Q_0$ (red arrow), as well as a pure \emph{DOSC} $h(r)$ are shown as a dashed curve (cyan) in the top panel. 
}
\label{fig:ln_rhr}
\end{figure}

The total correlation function $h(r)$ can be expressed in a more convenient form, compared to the usual $h(r)$ versus $r$ analysis, that better displays it's asymptotic behaviour, given by
\begin{equation}
    \ln(r|h(r)|)=-a_0r+\ln(B|\cos[Q_0r-\delta)]|),
\end{equation}
where the inverse screening lengths $a_0$, oscillation frequencies $Q_0$, and phases $\delta_{++}, \delta_{+-},\delta_{--}$ can be characterized. Figure~\ref{fig:ln_rhr} compares the results of MD and OZ/HNC (HNC for short) for the three ion-ion partials (the details for MD and HNC are provided in the SI). Several points need to be highlighted: (1) the dominant MD Er-Er correlations are well reproduced by HNC, (2) the MD Er-Cl and Cl-Cl correlations are not as well reproduced by HNC, (3) the MD simulations become noisier at large $r$ and at low concentrations (as shown in the SI) compared to HNC, and (4) the asymptotic behaviour of the partial waves displays spatial variations not described exactly by a \emph{single} pure \emph{DOCS} $h(r)$ partial wave with fixed $a_0, Q_0$ and $\delta$ as seen in the $h_{+-}$ and $h_{--}$ partial waves, however, they are close to the $d$-spacing in the $h_{++}$ partial wave. A few other observations about HNC and MD are also warranted. The MD simulations are more useful at higher concentrations \emph{e.g.,} at $c>1$ m, since there are more ions and the $a_0$ and $Q_0$ are more readily ascertained from the asymptotic behavior of $h(r)$. Unfortunately, the KT region is between $0<c<1$ m, where obtaining $a_0$ and $Q_0$ becomes problematic due to lack of asymptotic convergence and instead fitting the prepeak region of $h(Q)$ is the preferred route. Our view is that running larger and longer MD simulations doesn't really solve this issue, whereas carefully constructed HNC models can be used to obtain $a_0$ and $Q_0$ from asymptotically converged correlation functions.\cite{Warren2013,Gao-2023}

\begin{figure}[tbh]
\centering
\includegraphics[width=\columnwidth]{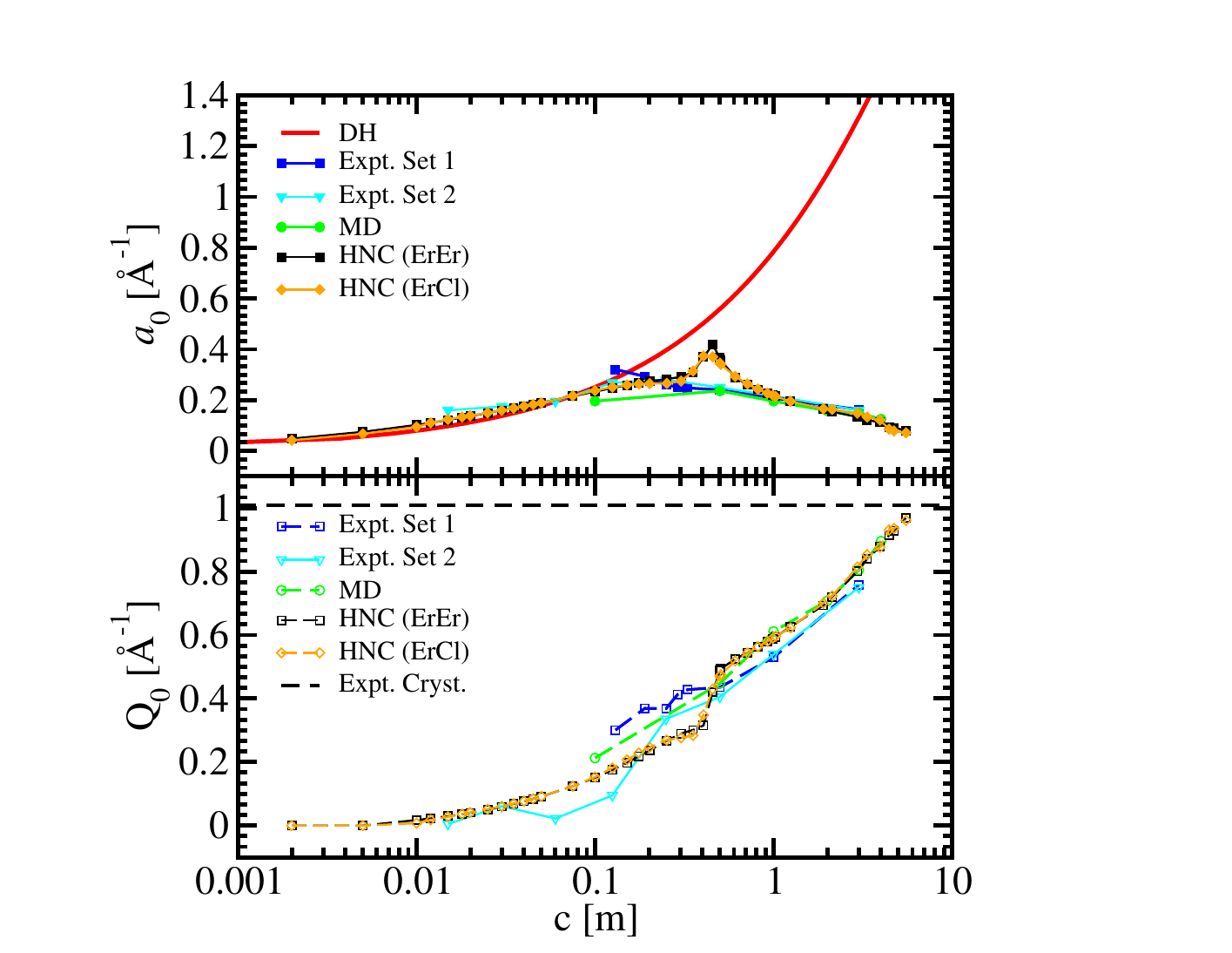}
\caption{A comparison of the $a_0$ and $Q_0$ from the experimental SAXS
spectra for aqueous \ce{ErCl_{3}}, MD simulation, HNC calculation,
and DH theory. In the bottom panel, a black dashed line denoting the $Q_0$ was found experimentally from the experimental (100) Bragg peak of the crystalline \ce{ErCl3} hexahydrate.}
\label{fig:ErCl3_aoQo}
\end{figure}

Using the \ce{ErCl3} SAXS data, MD simulations with all ten partials, and  three partials from HNC, we extract $a_0$ and $Q_0$ for a range of concentrations shown in Figure~\ref{fig:ErCl3_aoQo} by fitting $\hat{h}^{\rm TSF}(Q)$ to the prepeak X-ray signal $S_X(Q)$ generated from the underlying ten $\hat{h}_{ij}(Q)$ partials. The HNC long-range asymptotic \emph{DOSC} form for $h(r)$ was fit in $r$-space to the \ce{Er}-\ce{Er}/\ce{Er}-\ce{Cl} HNC total correlation functions to obtain $a_0$ and $Q_0$. The HNC calculations employ a mean field description of water where only the three ion-ion partials ($++$, $+-$, $--$) are used and the infinite dilution ion-ion potentials of mean force (PMFs) were calculated from MD simulations and used as input for the HNC calculations. The $a_0$ and $Q_0$ data shown in Figure~\ref{fig:ErCl3_aoQo} display a more complex behaviour than the idealized KT \emph{e.g.,} from Kjellander mentioned above (as well as other TEs shown in the SI). For \ce{ErCl3}, the SAXS as well as MD results show no pronounced cusp in $a_0$ at $c=0.45$ m whereas the HNC results do. At low concentrations $c<0.1$ m, $a_0$ converges to the DH behaviour. For $Q_0$, as the concentration is lowered, experiment, MD, and HNC consistently show a gradual approach to zero instead of the idealized KT behaviour of dropping to zero where the cusp in $a_0$ occurs. At the highest concentration near saturation, $Q_0$ converges to that for the experimental first Bragg peak of the crystalline \ce{ErCl3} hexahydrate\cite{Ivanov-1969}. These non-ideal KT \ce{ErCl3} results can perhaps be understood by considering that \ce{Er}$^{+3}$-\ce{Er}$^{+3}$ correlations are very strong (large Coulomb coupling) and hence oscillations persist to much lower concentrations\cite{Smirnov-2013,Soderhom-2009}. Furthermore, this means that oscillations persist even though the peak in $a_0$, minimum in screening length $\lambda$, has already occurred as the concentration is decreased. Again, one cannot generally assume that exponential to damped oscillatory decay occurs at the KT but instead the concentration dependence of $a_0$ and $Q_0$ must be considered independently.

\begin{figure}[tbh]
\centering
\includegraphics[width=\columnwidth]{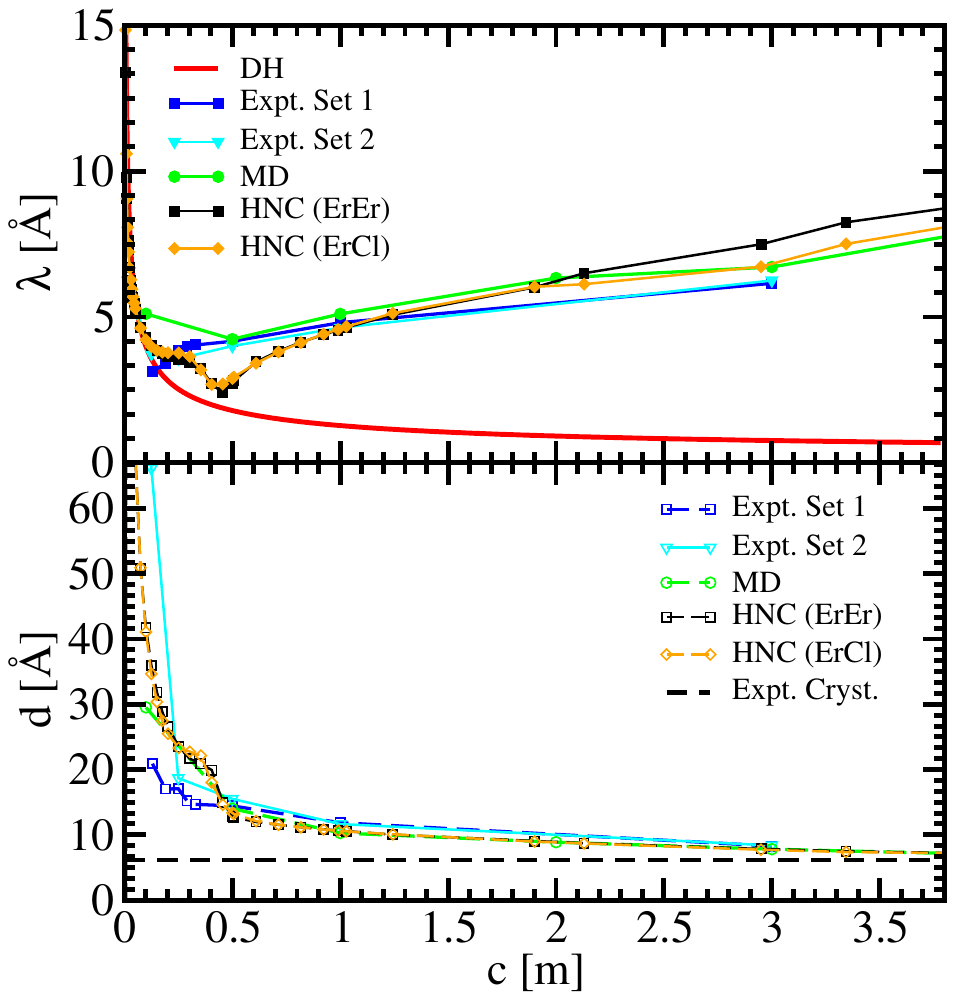}
\caption{A comparison of the $\lambda$ and $d$ from the experimental SAXS spectra for aqueous \ce{ErCl_{3}}, MD simulation, HNC calculation,
and DH theory. The experimental Bragg spacing for the (100) peak for the
crystalline \ce{ErCl3} hexahydrate is $d=6.22$\,\AA.}
\label{fig:ErCl3_lambda-d}
\end{figure}

For convenience, Figure~\ref{fig:ErCl3_lambda-d} shows the $r$-space screening lengths $\lambda$ and $d$-spacings corresponding to the data in Figure~\ref{fig:ErCl3_aoQo}. These comparisons show excellent agreement between the measured SAXS data and those from MD and HNC. These results represent the first joint experimental and molecular theory verification and validation of the non-Debye screening lengths and periodicities in concentrated aqueous \ce{ErCl3}. Importantly, the HNC results are able to capture the long-range asymptotics of ion-ion correlations where direct MD loses it's utility to yield consistent $a_0$ and $Q_0$ at $c<1$ m. 

To summarize what we have learned thus far, we find that our analysis of the \emph{asymptotic behaviour} of the ion-ion correlation functions in aqueous electrolytes yields accurate quantification of screening, periodicity, as well as the interpretation, prediction, and validation of both measured SAXS spectra and theory. Interestingly, the MD results (Fig. S10) show that at relatively high ion concentrations, the $Q_0$ values are almost consistent for the ion-ion, ion-water, and water-water correlations while at relatively low ion concentrations, the $Q_0$ values for ion-ion correlations tend to be smaller than ones for the ion-water and water-water correlations.

\begin{figure}[tbh]
\centering
\includegraphics[width=\columnwidth]{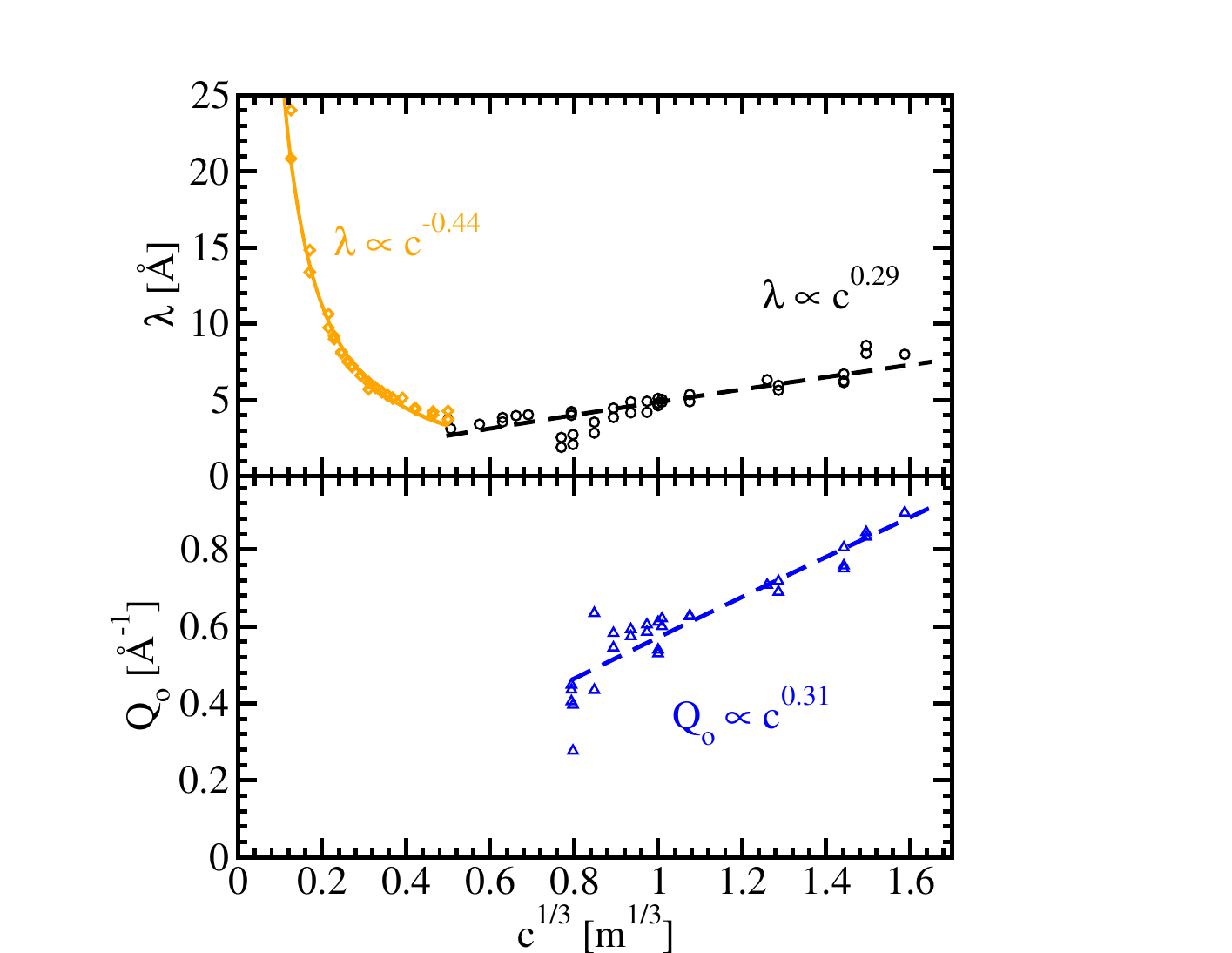}
\caption{Scaling relations in $\lambda$ and $Q_0$ as functions of \ce{ErCl3} concentration. All relevant data from SAXS, MD, and HNC have been included in the fitting. It is found that that $\lambda=1/a_0\varpropto c^{1/3}$ and $Q_0=2\pi/d\varpropto c^{1/3}$, \emph{i.e., quasi-cubic} scaling. The ratio of $\lambda/\lambda_{\rm D}\sim c^{5/6}$ for $c\geqslant 0.13\,m$.}
\label{fig:scaling_lambda_Qo}
\end{figure}

A question that naturally arises is: how do the non-DH screening lengths and oscillation frequencies scale with concentration? Figure~\ref{fig:scaling_lambda_Qo} shows the results of fitting the combined SAXS, MD and HNC data for $\lambda$ and $Q_0$ for aqueous \ce{ErCl3}. At low concentrations, the $\lambda\varpropto c^{-0.44}$ which is similar to the DH scaling $\lambda_{\rm D}\varpropto c^{-1/2}$. Above the KT region, a common quantity to report is how the screening changes with concentration expressed as the ratio of screening lengths $\lambda /\lambda_{\rm D}\varpropto c^{\alpha}$, where $\alpha$ is the scaling exponent. Our results give $\lambda\varpropto c^{\,0.29}\sim c^{1/3}$ \emph{i.e., quasi-cubic} scaling, yielding $\lambda /\lambda_{\rm D}\varpropto c^{5/6}$. Previous SFA measurements by Perkin \emph{et al.} found $\lambda /\lambda_{\rm D}\varpropto c^{3/2}$.\cite{SmithLeePerkin2016,LeeSmithPerkin2017} In contrast, using a charge-frustrated Ising model, Ludwig \emph{et al.} finds $\lambda /\lambda_{\rm D}\varpropto c$.\cite{Ludwig-2018} Rotenberg \emph{et al.}, using MD simulations of atomic and molecular ions, find a scaling range $\lambda /\lambda_{\rm D}\varpropto c^{1/2}$ to $c^{3/4}$.\cite{Rotenberg-2018,Rotenberg2018} For $Q_0$, we also find $Q_0\varpropto c^{\,0.31}\sim c^{1/3}$, equivalent to the \emph{quasi-cubic} scaling found in our previous work on \ce{CaCl2}\cite{Evgenii-2020} as well as that found in several other studies on aqueous electrolytes.

\begin{figure}[tbh]
\centering
\includegraphics[width=\columnwidth]{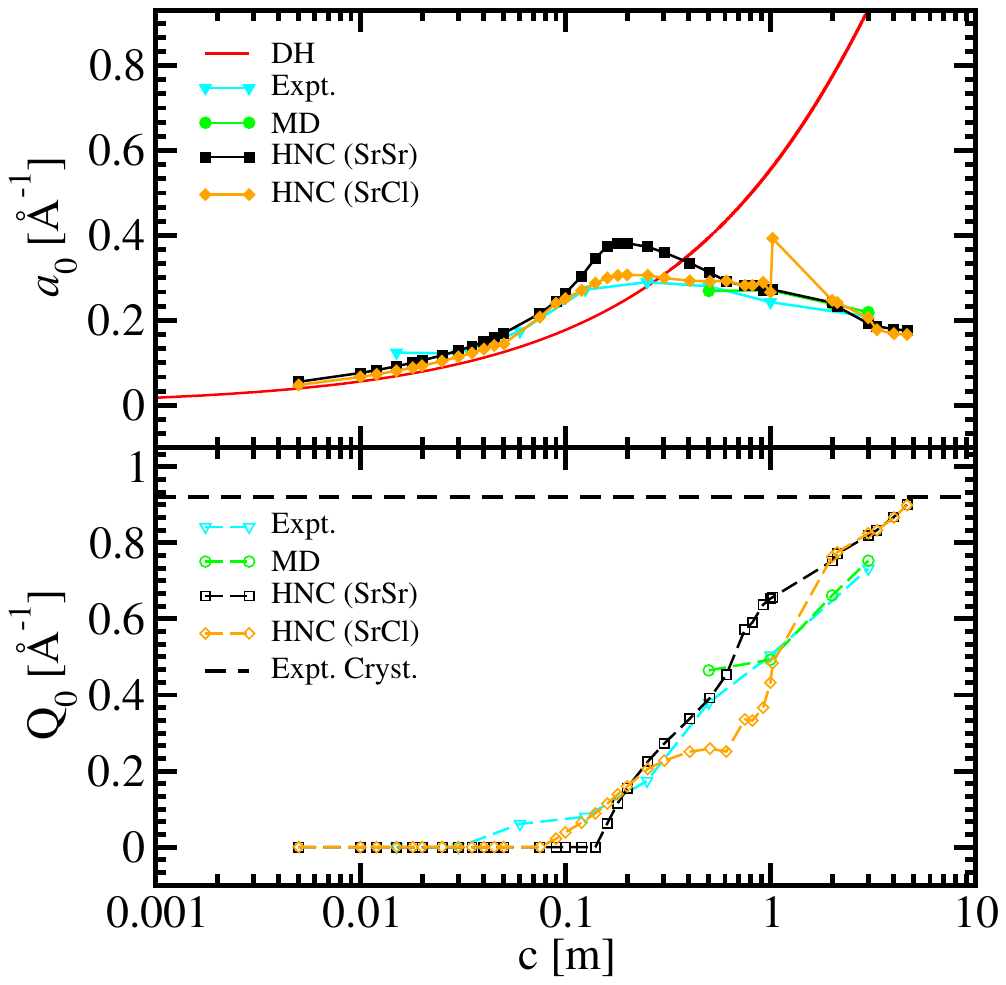}
\caption{A comparison of the $a_0$ and $Q_0$ from the experimental SAXS
spectra for aqueous \ce{SrCl_{2}}, MD simulation, HNC, and DH theory.}
\label{fig:SrCl2_2pan_a0-Q0}
\end{figure}

Figure~\ref{fig:SrCl2_2pan_a0-Q0} shows our $a_0$ and $Q_0$ results for aqueous \ce{SrCl2}. They are similar to \ce{ErCl3} but with the important difference due to the reduced Coulomb coupling in the lower valence \ce{Sr}$^{+2}$-\ce{Sr}$^{+2}$ correlations. In this case, the more standard KT behaviour is seen where $Q_0$ falls to zero around the the peak in $a_0$. We find very similar behaviour for aqueous \ce{CaCl2}, whose results are shown in the SI. It is worth noting that the fitting of both the experimental and simulation data are extremely sensitive and as such the uncertainties associated with $a_0$ and $Q_0$ values reported in the Figures can be about $0.1-0.2$ \AA$^{-1}$ depending on the concentration and the $r$-region or $Q$-region chosen for the fitting procedure.

For aqueous \ce{NaCl}, we don't find a pronounced SAXS prepeak - see SI. Instead, we find more of a broad spectral ``hump" at $Q\approx 0.7\,\text{\AA}^{-1}$ even at $c=6$ m \emph{i.e.,} very near saturation. The MD and HNC \ce{Na+}-\ce{Na+} correlations are smaller than the $+3$ and $+2$ cations. As mentioned previously, aqueous \ce{NaCl} is, as well as other alkali halides, notoriously difficult to extract information from the prepeak region with X-rays due to the compounding problems of low Coulomb coupling and low contrast. Calculating $S_X(Q)$ from the MD data doesn't yield the measured weak SAXS ``shoulder". However, we can still obtain $a_0$ and $Q_0$ from the asymptotic behavior of the \ce{Na+}-\ce{Na+} MD and HNC correlation functions consistent with those determined from fitting the SAXS ``hump" with $\hat{h}_{ij}^{\rm TSF}(Q)$, yielding $a_0=0.18\,\text{\AA}^{-1}$ and $Q_0=0.755\,\text{\AA}^{-1}$ which seem physically reasonable within the current context.

\subsubsection{Electrostatic Description of Electrolytes}

Screening of ions in an aqueous electrolyte is, by definition, an electric phenomenon, however, up to this point the screening and oscillation frequency have been determined from analyzing SAXS prepeaks, MD, and HNC. Let's consider the electric potential without the linearization and closure assumptions used in DH theory, where the response charge densities to a cation $z_+$ and anion $z_-$ are
\begin{equation}
\begin{split}
&\rho_{+}(r)=z_{+}en_{+}g_{++}(r)+z_{-}en_{-}g_{+-}(r)\\
&\rho_{-}(r)=z_{+}en_{+}g_{+-}(r)+z_{-}en_{-}g_{--}(r).
\end{split}
\end{equation}
The relevant differences in correlations are $\Delta h_{+}=h_{++}-h_{+-}$ and $\Delta h_{-}=h_{+-}-h_{--}$,
\begin{equation}
\begin{split}
&\Delta h_{+}=\frac{B^+e^{-a_0r}}{r}\left(\cos[Q_0r-\delta_1]-\cos[Q_0r-\delta_2]\right)\\
&\Delta h_{-}=\frac{B^-e^{-a_0r}}{r}\left(\cos[Q_0r-\delta_2]-\cos[Q_0r-\delta_3]\right)\\
\end{split}
\label{eqn_deltah}
\end{equation}
can be employed to obtain useful analytic expressions for comparison with MD and HNC. Here we compare the MD and HNC electric potentials and ion-atmospheres to analytic expressions assuming \textit{DOSC} ion-ion total correlation functions $h_{ij}(r)$ where the amplitude $B$, $a_0$, and $Q_0$ are the same but only their phases ($\delta_1=\delta_{++}$, $\delta_2=\delta_{+-}$, or $\delta_3=\delta_{--}$) differ.
Analyzing the electric potentials and ion atmospheres arising from the underlying ion-ion charge density correlations provides insight into the changes in electrolyte structure that give rise to the minimum in the screening length $\lambda$, oscillation frequency, and spatial extent. The minimum in screening length arises from the surrounding cations and anions as well as water orientations mitigating the spatial extent of the ion's potential and atmosphere.  
Stillinger's zeroth moment (a.k.a. electroneutrality condition) is given by
\begin{equation}
S_{0}^{\infty}=4\pi\int_{0}^{\infty}r^2\rho_{j}(r)dr=-z_{j}e,
\end{equation}
showing that at a sufficiently large distance from the central cation $z_j$, the electrolyte responds by generating the opposite charge.
We can investigate the spatial extent of the ion atmosphere required to achieve electroneutrality by evaluating
\begin{equation}
    S_{0}(r)=4\pi\int_{0}^{r}r'^2\rho_{j}(r')dr'
\end{equation}
whose asymptotic behaviour is given by
\begin{equation}
\begin{split}
&S_{0}(r)=-z_{j}e[a\,+\\
&e^{-a_0r}\left(b\,\cos[Q_0r-\delta_s]+c\,\sin[Q_0r-\delta_s]\right)],
\end{split}
\end{equation}
where $a, b,$ and $c$ are given in the SI. Using Poisson's equation for the asymptotic behaviour of response charge density, $\nabla^2\phi_j^{\rm resp}(r)=-4\pi\rho_j(r)$, and the zeroth moment condition (to obtain $B^+$ or $B^-$ in Eq. \ref{eqn_deltah}), yields the response electric potential caused by either cation $z_+$ or anion $z_-$,
\begin{equation}
\begin{split}
&\phi_{j}^{\rm resp}(r)=-\frac{z_{j}e}{\varepsilon r}e^{-a_0r}\\
&\left[\frac{\left(a_0^{2}-Q_0^{2}\right)\sin[Q_0r-\delta_{s}]+2a_0Q_0\cos[Q_0r-\delta_{s}]}{\left(a_0^{2}-Q_0^{2}\right)\sin[\delta_{s}]-2a_0Q_0\cos[\delta_{s}]}\right]\\
&-\frac{z_{j}e}{\varepsilon r},
\end{split}
\end{equation}
where $\delta_{s}=(\delta_{m}+\delta_{n})/2$ and $m$ and $n=$ 1 and 2 for cations and 2 and 3 for anions, respectively.
One can see that this response electric potential does not include the ion's self-electric potential $\phi_j^{\mathrm{self}}$. Thus, the total electric potential is $\phi_j^{\mathrm{tot}}=\phi_j^{\mathrm{resp}}+\phi_j^{\mathrm{self}}$. If we take the limit
\begin{equation}
\lim_{Q_{0}\rightarrow0}\phi_{j}^{\rm resp}(r)=-\frac{z_{j}e}{\varepsilon r}\left[1-e^{-a_{o}r}\right],   
\end{equation}
and add the ion's self-potential, one recovers the DH limit (where the ion size $a$ is small compared with $\lambda_{\rm D}$) for the total electric potential
\begin{equation}
\phi_{j}^{\rm tot}(r)=\frac{z_{j}ee^{-a_0r}}{\varepsilon r}.
\end{equation}

\begin{figure}[tbh]
\centering
\includegraphics[width=\columnwidth]{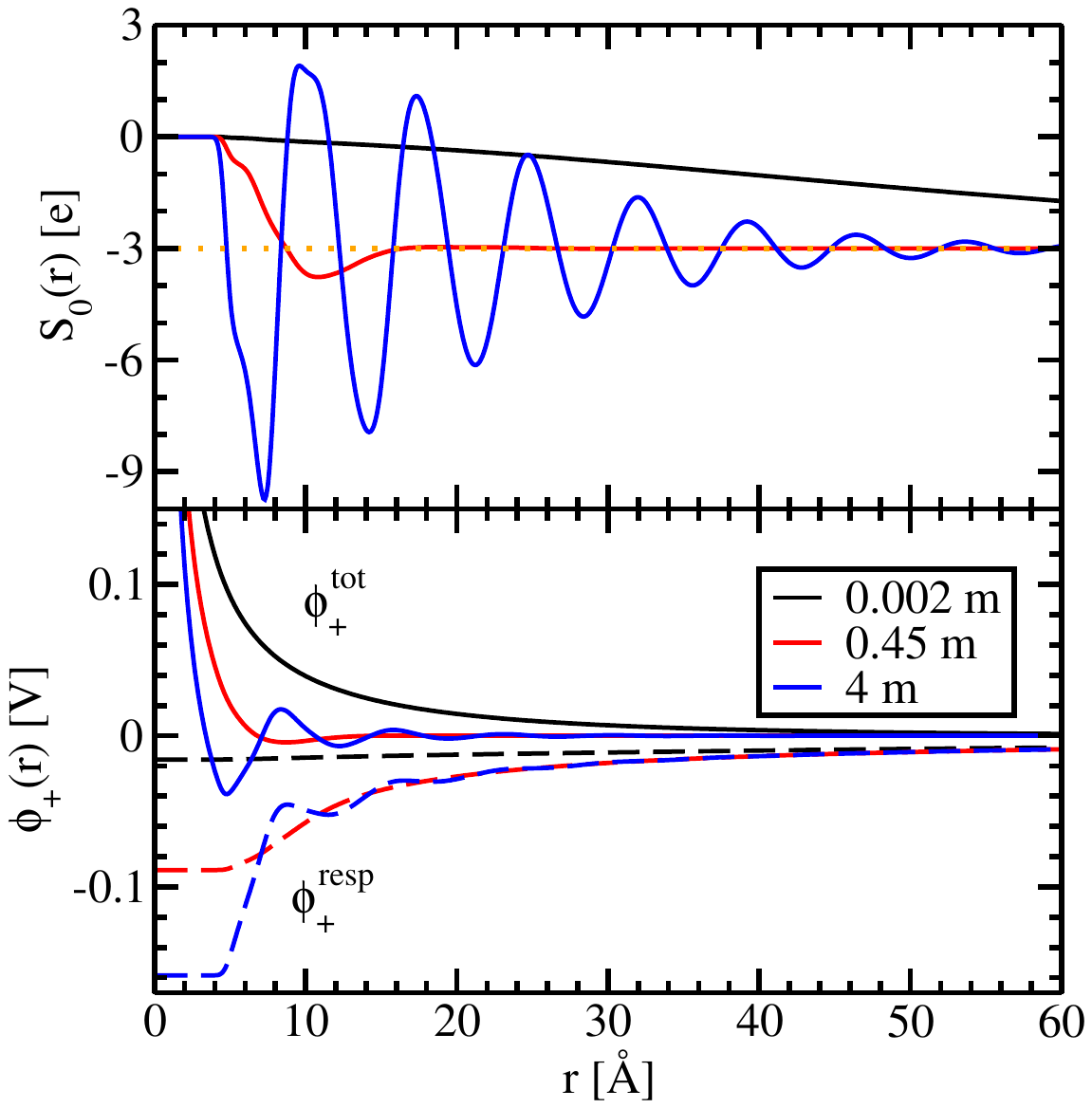}
\caption{\ce{Er}$^{+3}$ ion atmosphere $S_{0}(r)$ (top panel), response electric potential $\phi_{+}^{\mathrm{resp}}(r)$ (dashed curves - bottom panel), and total electric potential $\phi_{+}^{\rm tot}(r)$ (solid curves - bottom panel), from HNC calculations of \ce{ErCl3} at 4 \emph{m} with $\lambda = 9\,\text{\AA}$ and $d=7.2\,\text{\AA}$, 0.45 m for which $\lambda= 2\,\text{\AA}$ is a minimum and $d=15\,\text{\AA}$, and 0.002 \emph{m} with $\lambda = 25\,\text{\AA}$ and $d=\infty$. The asymptotic ion atmosphere value of $S_{0}^{\infty}=-3\,e$ is shown for reference as an orange dashed line.}
\label{fig:KT_ion_atmos_phi_2pan}
\end{figure}

Figure~\ref{fig:KT_ion_atmos_phi_2pan} shows $S_{0}(r)$, $\phi_{+}^{\rm resp}(r)$, and $\phi_{+}^{\rm tot}(r)$ for aqueous \ce{ErCl3} at various concentrations, including where $\lambda$ is a minimum using the HNC ion-ion correlations. The top panel shows the \ce{Er}$^{+3}$ atmosphere $S_{0}(r)$ charge oscillations compared to what is happening with the response $\phi_+^{\rm resp}(r)$ and total $\phi_+^{\rm tot}(r)$ electric potentials in the bottom panel. In the bottom panel, one can see that the response and total electric potentials differ by the inclusion of the self-electric potential of \ce{Er}$^{+3}$, as well as the lower plateau displayed by the response potential corresponding to the \ce{Er}$^{+3}$ radius. The analytic expressions for $S_{0}(r)$ and $\phi_+^{\rm resp}(r)$ derived above, using the asymptotic $DOSC$ form for $h_{ij}(r)$, are able to quantitatively reproduce the MD (or HNC) results as shown in the SI.
For $S_{0}(r)$ at $4$ m (blue curve), the large charge swings (e.g., -7, +5, -5, +4, -3, +3, -2, +2, -1, +1, ...) above and below the electroneutrality line (dotted orange) at $S_{0}^{\infty}=-3\,e$, do not decay to less than $\pm 1\,e$ until $r\approx40\,\text{\AA}$. Thus, the \ce{Er}$^{+3}$ causes the electrolyte to \emph{over-respond} to the negative more than to the positive. But, the HNC charge oscillation amplitudes of $S_{0}(r)$ are too large compared with the MD results by about $\pm2\,e$ as shown in the SI. We note that the MD results suffer from asymptotic convergence issues - see SI. The general result is that the MD ion atmosphere decays to electroneutrality at slightly shorter distances than HNC. The spatial extent of the MD ion atmospheres of several nanometers is consistent with the experimental coherence lengths $\xi$ previously discussed. At $c=0.45$ m, where the minimum screening length $\lambda=2$\,\AA\ occurs, the $\phi_+^{\rm tot}(r)$ falls to zero most rapidly (though there are still oscillations since $Q_0\neq0$) and the ion atmosphere extends to about 18 \AA. At $c=0.002$ m one can see that $\phi_+^{\rm tot}(r)$ becomes even longer ranged than at $c=4$ m, with $\lambda=25$\,\AA. These results show how the long-range behavior, encoded in the $a_0(c)$ and $Q_0(c)$ probed by SAXS, are manifested in the underlying electric potentials and ion atmospheres. These results show that by modestly changing the electrolyte concentration, the character and spatial extent of the ion atmospheres change dramatically.

\begin{figure}[tbh]
\centering
\includegraphics[width=14cm,height=8cm,keepaspectratio]{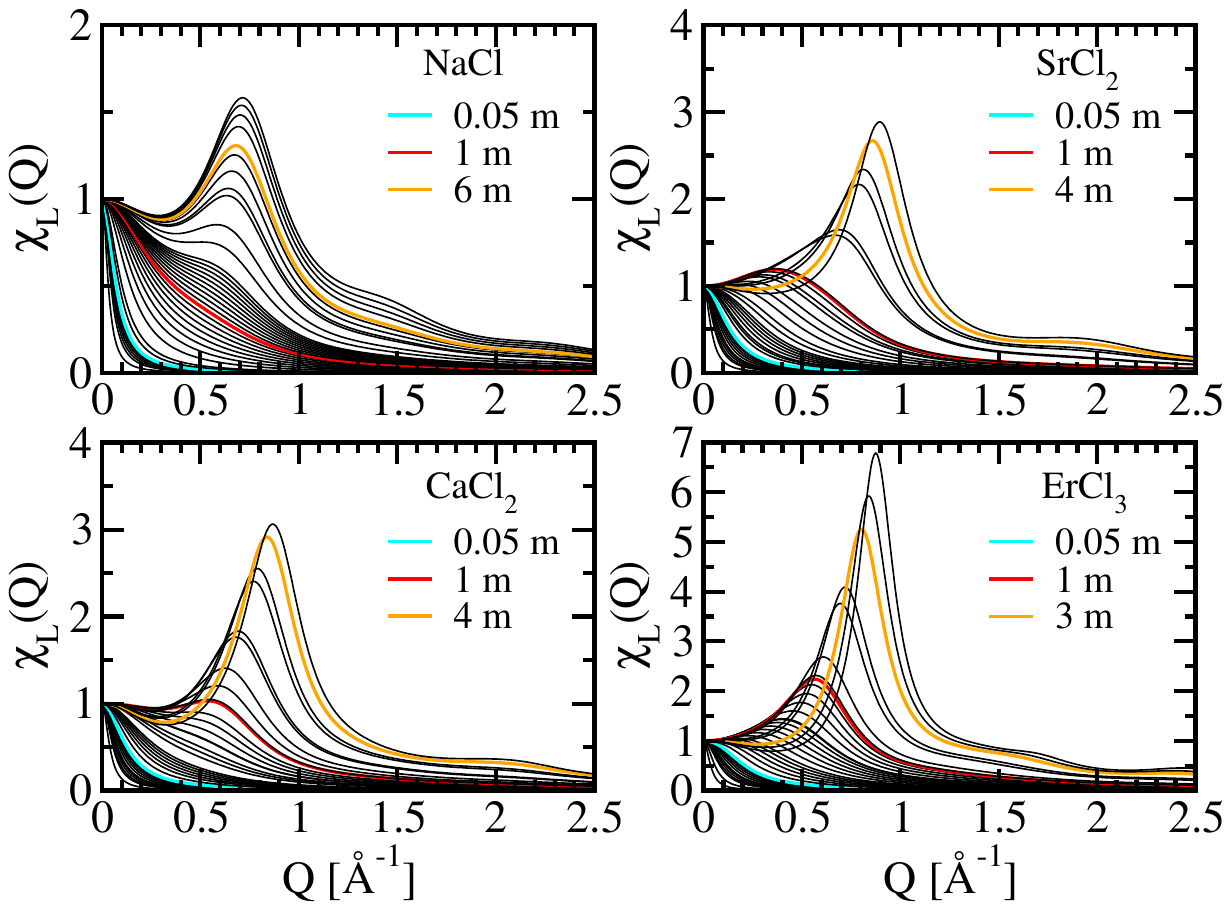}
\caption{Response dielectric susceptibility $\chi_{\rm L}(Q)$ from HNC calculations of all electrolytes studied in this work at various concentrations. Three intermdediate concentrations for each electrolytes are shown by cyan, red, and orange lines.}
\label{fig:sample_chi}
\end{figure}
A closely related property that can shed light on the screening behavior of electrolytes is the dielectric susceptibility.
Following the dielectric response theories for bulk materials, the response charge density can be characterized by the polarization response of the bulk electrolyte, $\mathbf{P}$, due to an external electric field, $\mathbf{E}$, which is related to the electric susceptibility $\chi$ or dielectric constant $\varepsilon$. In linear response in $Q$-space, $\mathbf{P}(Q)$ due to an external electric field $\mathbf{E}(Q)$ is related to the susceptibility via $\mathbf{P}(Q)=\chi(Q)\mathbf{E}(Q)$, in which $\chi(Q)$ is a scalar function for isotropic materials and contains the spatial information about the screening behavior of electrolytes. The longitudinal non-local dielectric constant $\varepsilon_{\rm L}(Q)$ and response dielectric susceptibility $\chi_{\rm L}(Q)$ are related by $\varepsilon_{\rm L}(Q)=1/(1-\chi_{\rm L}(Q))$. This dielectric response $\chi_{L}(Q)$ can be obtained through the fluctuation-dissipation theorem in terms of the ion-ion structure factor via $\chi_{\rm L}(Q)= \kappa^2_{\rm D}\zeta^{-1} S_{zz}(Q)Q^{-2}$ where 
\begin{equation}
S_{zz}(Q)=\sum_{i,j}z_{i}z_{j}S_{ij}(Q),
\end{equation}
with $\zeta$ is the ionic strength $\zeta = \sum_i z_i ^2 n_i$.  
The $Q$ and concentration-dependent $\chi_{\rm L}(Q,c)$ of the electrolytes we study are presented in Fig. \ref{fig:sample_chi}. Such response functions, $\chi_{\rm L}(Q,c)$, encode variations in spatial resonance or dispersion of the electrolytes response to an external field and thus provide a means to exploit the electrolyte's spatial response by tuning the concentration. At the highest concentrations of ErCl$_3$, SrCl$_2$, and CaCl$_2$, we see a resonance peak close to $Q\sim0.9$\,\AA$^{-1}$ while for NaCl, the resonance peak appears about $\sim0.7$ \AA$^{-1}$. As the concentration is lowered the evolution of the peak in $\chi_{\rm L}(Q)$ is consistent with the long-range behavior discussed above. We note, however, that the HNC results for $\chi_{\rm L}(Q)$ were obtained using only the mean field ion-ion partials compared to all ten partials. Thus, at low concentrations, the HNC $\chi_{\rm L}(Q)$ will not converge to that for pure water which is peaked around $Q\sim3$\,\AA$^{-1}$ but note that at $Q\sim0.9$\,\AA$^{-1}$ the 
nonlocal dielectric constant of pure water drops to $20$ (a wavelength which is comparable to the Bjerrum length of a monovalent ion, $\sim7$ \AA).\cite{Sutmann-1996} 
This implies that as ions are added to water a low-$Q$ long-range resonance appears in the electric susceptibility and increases in intensity as it approaches saturation.
On the nanometer scale, the non-local dielectric response, $\varepsilon(Q)$, arises from the behavior of ions and the interactions between them.\cite{Maggs2006} 
The field energy of a homogeneous fluid is proportional to $\sum_Q \varepsilon(Q) \mathbf{E}(Q)^2$; therefore,
at higher concentrations, we anticipate that the interactions are significantly influenced while at low concentrations such influences are less dramatic.

\begin{figure}[tbh]
\centering
\includegraphics[width=\columnwidth]{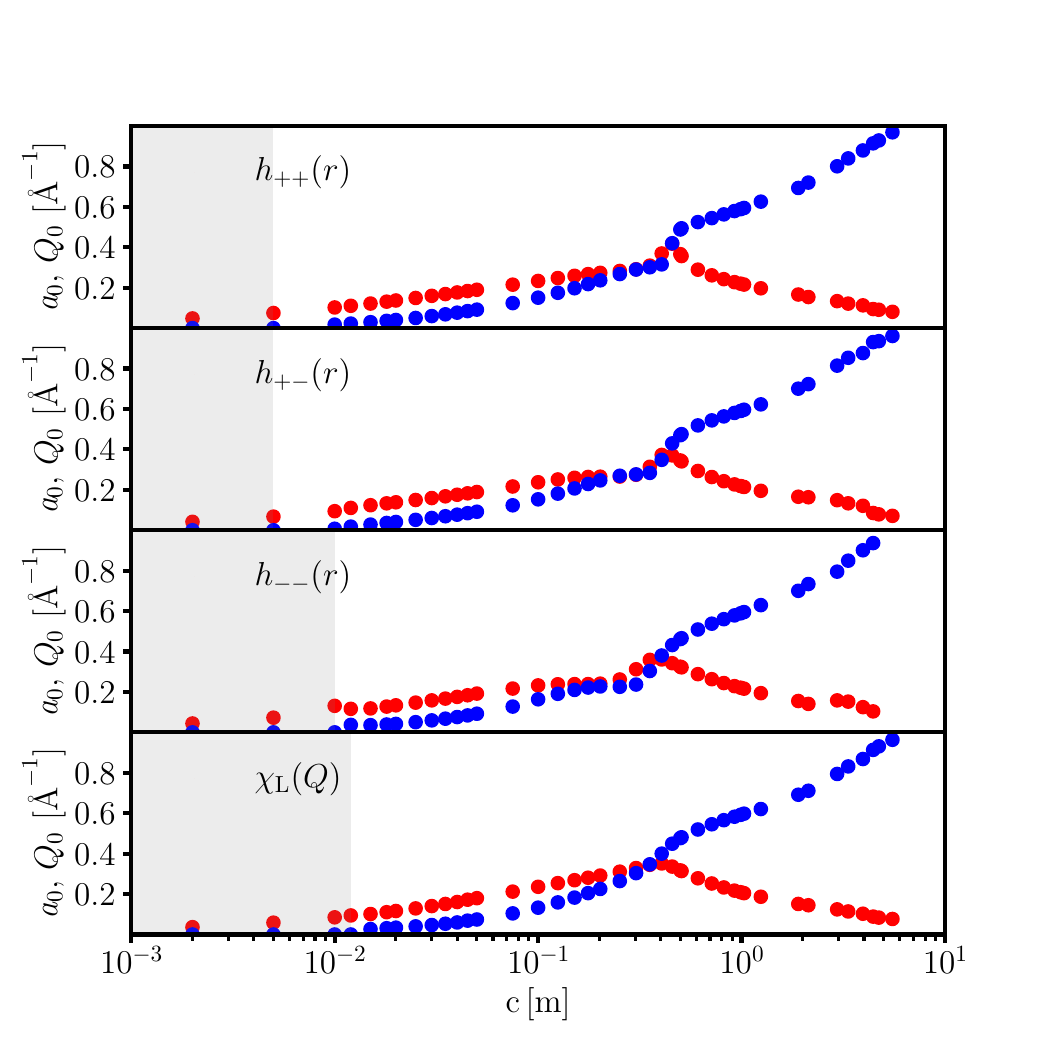}
\caption{Determination of $a_0$ and $Q_0$ for ErCl$_3$ in HNC calculations using three different partials as well as $\chi_{\mathrm{L}}(Q)$ (using Eq. \ref{gen_chi}). The shaded area shows where $Q_0 \rightarrow 0$.}
\label{fig:ErCl3_a0Q0_h_chi}
\end{figure}

Similar to the analysis of $I(Q)$ from which one can extract useful information about the correlations in the electrolyte systems, one can extract important information about the general screening behavior of electrolyte solutions in terms of $a_0$ and $Q_0$ at various concentrations and resolutions. Here we use a field theory approach, which is a simple way to study the behavior and properties of electrolytes at a microscopic level and
involves statistical averaging of many degrees of freedom to obtain a functional describing the system.\cite{Chandler-1993,Caillol-2003,Kardar-2007} The details will be presented in future work, but generally one can extract the pole structure from $\chi_{\rm L}(Q)$ as 

\begin{equation}
 \chi_\mathrm{L}(Q) = \frac{\kappa_s^2}{Q^2 + \kappa_s^2 - Q^2 \tilde{\Sigma}(Q)}
 \label{gen_chi}
\end{equation}
with the self-energy term given by
\begin{equation}
\tilde{\Sigma}^(Q)= \frac{A^2}{1+g(Q)} 
 \label{aux_f}
\end{equation}
where $\displaystyle g(Q)=\sum_{i=1} b_i Q^{2i}$ and $\kappa_s$ as the effective Debye parameter and the coefficients $A$, $b_i$ and $\kappa_s$ are fitting parameters.
Figure \ref{fig:ErCl3_a0Q0_h_chi} shows the pole structure for ErCl$_3$ from the total correlation functions of $h_{++}(r)$, $h_{+-}(r)$, and $h_{--}(r)$ as well as $\chi_{\mathrm{L}}(Q)$ extracted from Eq. \ref{gen_chi}. As can be seen, they are in good agreement (the pole structure comparison from such analyses will be presented in future work). 
Therefore, the response functions, $\chi_{\rm L}(Q)$, have the same single pole structure as that found from the asymptotic behavior of $h_{++}(r)$, $h_{+-}(r)$ and $h_{--}(r)$, as well as the measured $I(Q)$.
Interestingly, the behaviors of $\chi_{\mathrm{L}}(Q)$ can also be captured with TS distributions once the $Q$-region for the analysis is chosen carefully as discussed in the SI. Equation \ref{gen_chi}, however, can accurately represent the behavior of $\chi_{\rm L}(Q)$ for the studied electrolytes over a wider range in $Q$-space.

\section{Conclusion}

Ion-ion correlations, screening lengths, and oscillation frequencies
in various concentrated electrolyes have been investigated using synchrotron
SAXS, theory, MD, and HNC. From the SAXS measurements in the low-$Q$ prepeak
region of the scattering spectra we are able to extract screening
lengths $\lambda$ and inverse periodicities $Q_0=2\pi/d$, where
$d$ is the spatial correlation length, that agree exceedingly well with the MD and HNC
results. This approach allowed the Kirkwood transition to be investigated and quantified for various electrolytes.
The main conclusions of this work are: (1) at low concentration the screening length converges to the Debye H\"{u}ckel limit $\lambda_{\rm D}$, whereas the screening lengths $\lambda$ are appreciably larger than $\lambda_{\rm D}$ as the concentration
$c$ is increased with a scaling $\lambda/\lambda_{\rm D} \sim c^{5/6}$, (2) that the $d=2\pi/Q_0$-spacing approach the experimental crystalline Bragg spacing at concentrations near saturation, (3) we derive an analytic form for an ion's electric potential and ion atmosphere in good agreement with the exact MD results, (4) the non-local $\chi(Q)$ suggests that at higher electrolyte concentrations where the cross-over decay occurs, the interactions are mediated by a spatially varying dielectric while at low concentrations the interactions are mainly screened by the bulk dielectric constant, and (5) we show that a single cation-cation pole describes the concentration dependence of $S_X(Q)$, $S_{zz}(Q)$, and $\chi_{\rm L}(Q)$.  

Finally, SAXS provides a direct connection to the ion-ion correlations, as well as quantities derived from those correlations, than those inferred from SFA which are based on the detection of the interaction between two planar charged surfaces.\cite{Israelachvili-1988,Evans-1999,Liang2007} Ultimately, a single pole involving the cation-cation correlations are able to capture quantitatively the essential features of the electrolyte response as measured by SAXS. These important length scales provide a pathway to describe electrolyte phenomena beyond using the DH theory.

\section{SUPPLEMENTARY INFORMATION}
The experimental and computational details are presented in the supplementary information (SI). 

\section{ACKNOWLEDGMENTS}
The authors would like to thank Jan Ilavsky at the Advanced Photon Source, Argonne National Laboratory, and Jaehun Chun at the Pacific Northwest National
Laboratory (PNNL) for helpful discussions. The authors were supported by the U.S. Department of Energy (DOE),
Office of Science, Office of Basic Energy Sciences, Division of Chemical
Sciences, Geosciences and Biosciences. PNNL is operated by Battelle for the U.S. DOE under Contract
No. DE- AC05-76RL01830. Computing resources were generously allocated
by PNNL\textquoteright s Institutional Computing program. DOE Contract
No. DE- AC02-06CH11357 supports the operation of the Advanced Photon Source
at Argonne National Laboratory.

\section*{REFERENCES}

\end{document}


\title{Supplementary Information: The First Direct Detection of Kirkwood Transitions in Concentrated Aqueous Electrolytes using Small Angle X-ray Scattering}

\author{Mohammadhasan  Dinpajooh}
\affiliation
{Physical and Computational Sciences Directorate, Pacific Northwest National Laboratory, Richland WA}
\author{Elisa  Biasin}
\affiliation
{Physical and Computational Sciences Directorate, Pacific Northwest National Laboratory, Richland WA}
\author{Christopher J. Mundy}
\affiliation
{Physical and Computational Sciences Directorate, Pacific Northwest National Laboratory, Richland WA}
\affiliation
{Department of Chemical Engineering, University of Washington, Seattle WA}
\author{Gregory K. Schenter}
\affiliation
{Physical and Computational Sciences Directorate, Pacific Northwest National Laboratory, Richland WA}
\author{Emily T. Nienhuis}
\affiliation
{Physical and Computational Sciences Directorate, Pacific Northwest National Laboratory, Richland WA}
\author{Sebastian T. Mergelsberg}
\affiliation
{Physical and Computational Sciences Directorate, Pacific Northwest National Laboratory, Richland WA}
\author{Chris J. Benmore}
\affiliation
{Advanced Photon Source, Argonne National Laboratory, Chicago IL}
\author{John Fulton}
\affiliation
{Physical and Computational Sciences Directorate, Pacific Northwest National Laboratory, Richland WA}
\author{Shawn M. Kathmann}
\affiliation
{Physical and Computational Sciences Directorate, Pacific Northwest National Laboratory, Richland WA}

\date{\today}

\maketitle

\section{Experimental Details}

The SAXS measurements were performed using three different beamlines, 6-ID-D, 12-ID-B and 9-ID-D at the Advanced Photon Source (Argonne National Laboratory, USA) in order to cover a very broad range of Q values ($0.005$ to $3.0$ \AA$^{-1}$) with the best possible S/N using different optical configurations and energies.
Beamline 6-ID-D provided a flux of 5$\times$1011 photons per second at an energy of  100 keV with beamsize of  500 (H)  $\times$ 500 (W) $\mu$m. A Varex CT4343 a-Si area detector was used to cover a range from $0.2 < Q < 3.0$ \AA$^{-1}$.
Beamline 12-ID-B provided a flux of 1012 photons per second at an energy of 13 keV with a beamsize of 200 (H)  $\times$  40 (W) $\mu$m. A 2 M-pixel Pilatus was used to cover a range from $0.03 < Q < 0.9$  \AA$^{-1}$.
Beamline 9-ID-D provided a flux of 1012 photons per second at an energy of 21 keV with a beamsize of 200 (H)  $\times$ 100 (W) $\mu$m beam. This USAXS beamline uses a Bonse-Hart configuration that was set to cover a range from $0.005 < Q < 0.025$  \AA$^{-1}$.\cite{Ilavsky2018}

 Samples were sealed in borosilicate capillaries having nominal diameters of $1$ or $1.5$  mm OD and wall thicknesses of approximately $10$ $\mu$m. The measurements were performed at room temperature. For the SrCl$_2$ and ErCl$_3$ samples, that were run at beamline 12-ID-B, the capillaries were further hand-selected to produce sets of capillaries having diameter variability of  less than $\pm 50$ $\mu$m in order to provide the best possible uniformity for comparisons of the total scattered intensity and for cases requiring background subtraction of pure water spectra. The ErCl$_3$ samples were prepare at pH 1.5 using an HCl stock solution in order to avoid hydrolysis. Scattering patterns were radially integrated using the FIT2D package\cite{Hammersley2016} while other corrections were applied using custom software. The background removal and data treatment follows methods previously described.\cite{Skinner2013,Skinner2016}

A non-linear least square fitting method was used to fit the Teubner-Strey functional form to the experimental spectra using the parameters of $Q_o$, $a_o$, and $\delta$, representing the peak position, the peak width and frequency phase shifts, respectively, as defined within the Teubner-Strey function. A linear background corrections ($m Q + b$) was included in the fitting of the I(Q) signals.

\begin{figure}[tbh]
\centering
\includegraphics[width=0.9\textwidth]{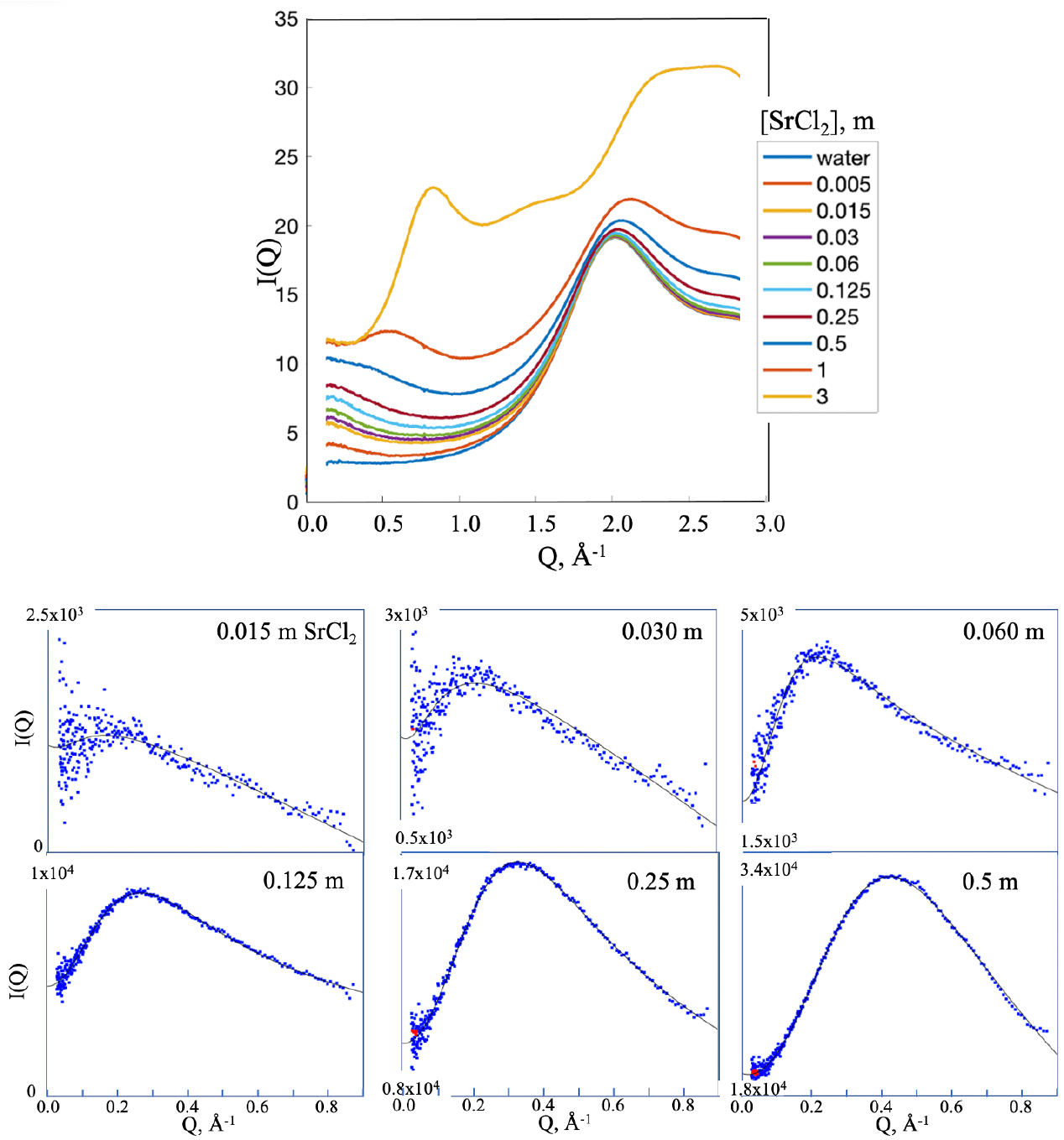}
\caption{Experimental I(Q) spectra for a series of aqueous SrCl$_2$ solutions acquired at beamline 6-ID-D (top panel).   Experimental I(Q) spectra with fits to the Teubner-Strey distribution for a series of aqueous SrCl$_2$  solutions acquired at beamline 12-ID-B (bottom panel). 
}
\label{compare_Oct_caps}
\end{figure}

\begin{figure}[tbh]
\centering
\includegraphics[width=0.9\textwidth]{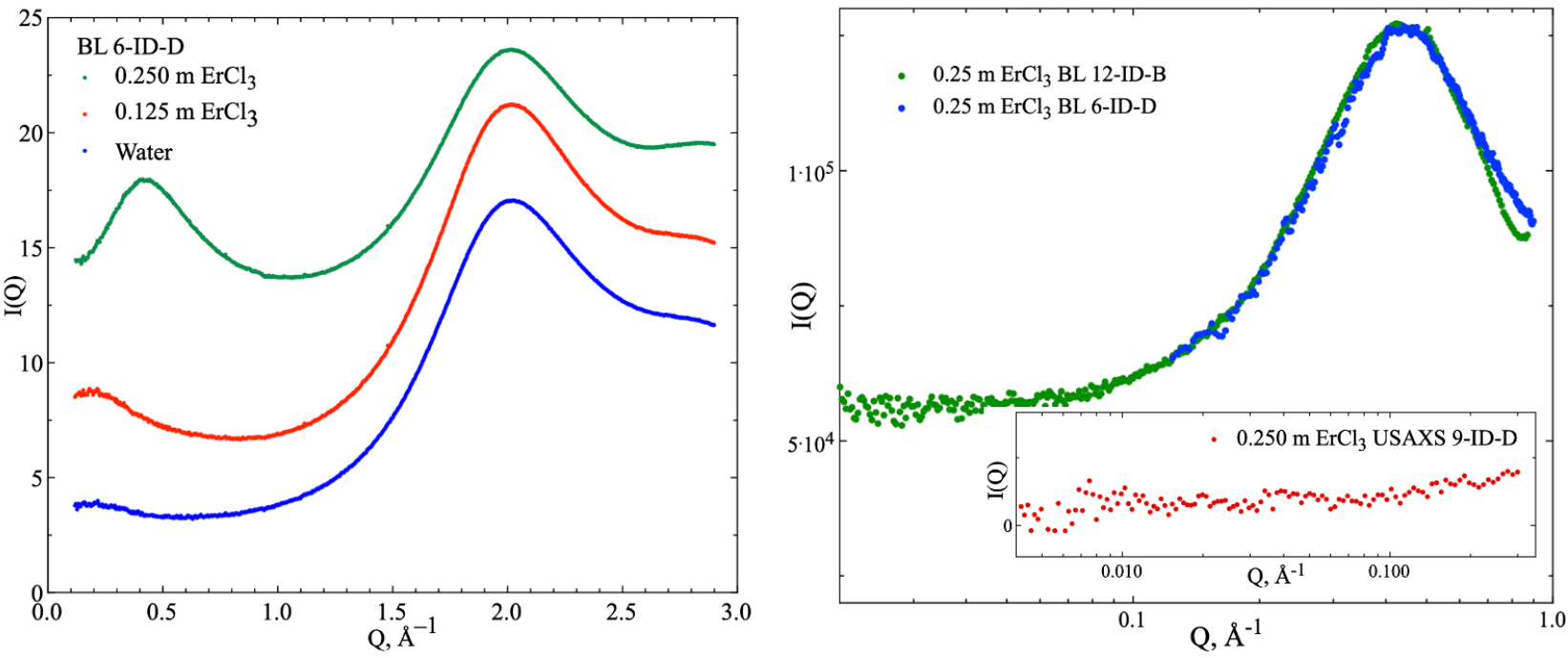}
\caption{ Experimental I(Q) spectra (left panel) for a series of aqueous ErCl$_3$, pH 1.5 solutions acquired at beamline 6-ID-D .  Experimental I(Q) spectra (right panel) for  0.25 m ErCl$_3$, pH 1.5 acquired at three different APS beamlines demonstrating the equivalency of the spectra and the absence of addition structure in the region from $0.004 < Q < 0.1$ \AA$^{-1}$.  For these spectra the pure water background has been subtracted and correction for sample holder and air scattering have been made.  In all cases I(Q) is in arbitrary units. 
}
\label{compare_beamlines}
\end{figure}

\newpage
\clearpage

\begin{figure}[tbh]
\centering
\includegraphics[width=\textwidth]{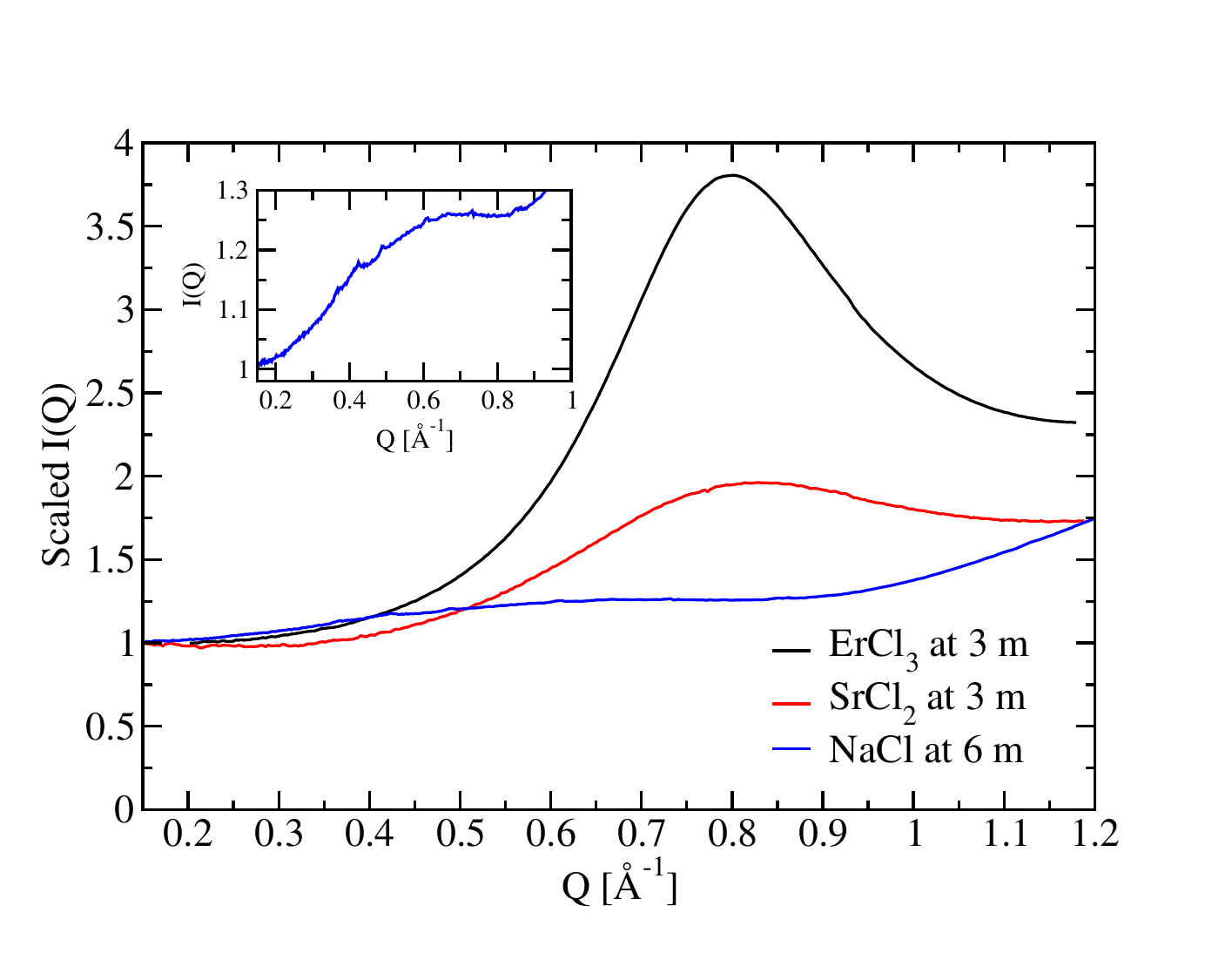}
\caption{ The effect of electron density contrast and valence on the pre-peak SAXS signals for ErCl$_3$, SrCl$_2$, and NaCl electrolyte solutions at high concentrations. The inset focuses on the pre-peak SAXS signal for NaCl electrolyte solution at 6 m. 
}
\label{compare_SAXSexpt}
\end{figure}

Figure \ref{compare_SAXSexpt} compares the pre-peak SAXS signals for ErCl$_3$, SrCl$_2$, and NaCl electrolyte solutions at high concentrations and shows the effect of electron density contrast and valence in the SAXS signal. As can be seen, when the cation is Er$^{3+}$, the X-ray signal is significantly strong and as the electron density and the valence on the cation decrease, the signal relative intensity considerably decreases such that the pre-peak for NaCl at 6 m is vaguely visible.  
Experimental I(Q) spectra for 6 m aqueous NaCl, acquired at APS beamline 12-ID-B. One can see that even at the highest concentration of 6 m only a ``hump" at $Q \sim 0.7$ \AA$^{-1}$ is observed. These spectra have had the capillary background subtracted. In all cases, I(Q) is in arbitrary units.

\clearpage



\section{Computational Details}

\subsection{Molecular dynamics simulations}

The molecular dynamics (MD) simulations were performed using the LAMMPS software program\cite{Plimpton1995} at $298$ K and a pressure of $1$ atm.  
The periodic boundary conditions were applied in 3-dimensions and the MD simulations were performed in the isothermal-isobaric ensemble (NPT) using the Nos$\rm{\acute{e}}$-Hoover thermostat and barostat.
The standard velocity-Verlet time integrator was used with a time step of $2$ fs and the SHAKE procedure was used to conserve the intramolecular constraints.
The Lennard Jones (LJ) and real-space part of the Coulombic interactions were truncated at $10$ \AA\ with additional switching/shifting functions that ramped the energy smoothly to zero between $10$ and $14$ \AA.
The conducting metal (tinfoil) boundary conditions were used to treat the electrostatic interactions and the particle-particle particle-mesh solver was used.
In the initial setup, at least $8000$ water molecules were initially placed in cubic lattices and the ions were then deposited between water molecules using the LAMMPS fix deposit command by avoiding the overlaps between the ions and water molecules considering the periodic boundary conditions.
The initial setups were then equilibrated for about $2-5$ ns and the production MD simulations were performed for $20$ ns, where $50000$ configurations were used to obtain the structural properties such as the radial distribution functions.

The SPC/E water model was used and the force fields for the ions used in this work are presented in Table \ref{pot_param}.
Following the potential energy model developed by Ribeiro for EuCl$_3$,\cite{Ribeiro2006} the current available potential parameters for EuCl$_3$ and ErCl$_3$ \cite{Merz2014} were used to scale the Eu$^{+3}$ LJ  Ribeiro parameters in order to get the Er$^{+3}$ LJ parameters for this work. The SD model\cite{SD-1995} was used for Cl$^{-}$ LJ parameters when used in ErCl$_3$ solutions. 
For SrCl$_2$ and CaCl$_2$, the Kirkwood-Buff force fields (KBFFs) were used.\cite{KBF-NaCl,KBF-M2}

\begin{table}[tbh]
\centering
\caption{Force field (FF) parameters used for the ions and water molecules in the MD and HNC calculations. The Cl$^-$ FF parameters of the Smith-Dang (SD) model was used for the anions in ErCl$_3$ and NaCl electrolyte solutions. The Kirkwood-Buff FF (KBFF) parameters were used for CaCl$_2$ and SrCl$_2$ electrolyte solutions. The Lorentz-Berthelot rule was used for the non-bonded LJ potentials.}
\begin{tabular}{ccccc}
    FF & Site & $\epsilon$(kcal mol$^{-1}$) & $\sigma$(\AA) & $q$(e)  \\
     \hline
 SPC/E   & O & $0.1554$ & $3.166$ & $-0.8476$ \\
 SPC/E   & H & $0$ & $0$ & $+0.4238$ \\
 this work   & Er$^{3+}$ & $2.7127$  & $1.696$  & $+3$   \\
 SD    & Cl$^{-}$ & $0.1000$  & $4.40$ &   $-1$   \\
 SD    & Na$^{+}$ & $0.1300$  & $2.35$ &   $+1$   \\
 KBFF   & Sr$^{2+}$ & $0.1195$  & $3.10$  & $+2$   \\
 KBFF   & Ca$^{2+}$ & $0.1123$  & $2.90$ &   $+2$   \\
 KBFF   & Cl$^{-}$ & $0.1123$  & $4.40$ &  $-1$   \\
     \hline
\end{tabular}
\label{pot_param}
\end{table}

\clearpage
\subsection{Ornstein-Zernike Solver for Electrolyte Solutions}

The ion-ion correlations for the electrolyte solutions, in this work, were obtained by solving the Ornstein-Zernike (OZ) for a mixture of two ions and treating the water as a continuum with the dielectric constant of $80$ ($\epsilon_w = 80$). This, in turn, can significantly simplify the OZ equations for electrolyte solutions to get the ion-ion correlations, but requires accurate effective ion-ion interactions. The effective interaction potentials between the ions (mean-field interactions) were obtained directly from the MD simulations by calculating
the potential of mean forces (PMFs) between the ions in the dilute limit directly  of two ions in the explicit SPC/E water molecules (see Table \ref{pot_param}).
However, the long-range parts of the effective interactions cannot be extracted from MD simulations due to the limitations/uncertainties in the MD simulations. Therefore, it is assumed that the MD effective interactions can be split to the short-range, $u^{\mathrm{SR}}_{ij}(r)$, and long-range electrostatics contributions, $u^{\mathrm{LR}}_{ij}(r) $, as 
\begin{equation}
W^{\mathrm{MD}}_{ij}(r) = u^{\mathrm{SR}}_{ij}(r) + u^{\mathrm{LR}}_{ij}(r) 
\end{equation}

Interestingly, such a decomposition is very useful for solving the OZ equation numerically because once the short-range effective interactions between the ions are determined from the MD simulations, one can reasonably approximate the effective long-range electrostatics interactions between ions and improve the convergence.\cite{Ng-1974,Warren2013} In this work, we assumed that the short-range effective interactions between the ions can be extracted from MD simulations by simply subtracting the $z_i z_j/(\epsilon_w r)$ terms corresponding to ions with charges of $z_i$ and $z_j$    
from the full effective interactions between the ions from MD simulations (MD PMFs, $W^{\mathrm{MD}}_{ij}$):
\begin{equation}
u^{\mathrm{SR}}_{ij}(r) = W^{\mathrm{MD}}_{ij}(r) - \frac{z_i z_j}{\epsilon_w r}
\label{sr-u}
\end{equation}
where $W^{\mathrm{MD}}_{ij}(r)$ shows the asymptotic behaviors of the long-range electrostatics interactions; therefore, the raw MD PMFs obtained either from the thermodynamics integration or WHAM methods were shifted vertically to match $z_i z_j/(\epsilon_w r)$ at a reasonable $r$.  
It would be an interesting future project to extract $W^{\mathrm{MD}}_{ij}(r)$ from other methods.

The OZ equations were then solved using the hyper-netted chain (HNC) closure.
We used a modified version of SunlightHNC code (PNNL-SunlightHNC),\cite{Warren2013}
where the inputs to the PNNL-SunlightHNC code were the short-range effective interactions between ions (see Eq. \ref{sr-u}), ion charges, and ion concentrations.
The long-range electrostatic interactions in the PNNL-SunlightHNC code were treated as

\begin{equation}
u^{\mathrm{LR}}_{ij}(r) = k_{\mathrm{B}}T z_i z_j l_\mathrm{B} \frac{\erf \left[ r/(2\sigma) \right]}{r}
\end{equation}
where $l_\mathrm{B}$ is the Bjerrum length setting the magnitude of the long-range electrostatics interaction at a given temperature $T$ ($l_\mathrm{B} \approx 7$ \AA\ for water at room temperature), $ k_{\mathrm{B}}$ is the Bolzmann constant, $\sigma$ is the size (width) of the Gaussian charges. 
A $\sigma$ value of $1$ \AA\ was used in the calculations; therefore, at distances above 4 \AA, $u^{\mathrm{LR}}_{ij}(r) \approx  z_i z_j/(\epsilon r)$, which justifies the use of Eq. \ref{sr-u} to extract the short-range mean-field ion-ion potentials for the HNC code.
We also checked that consistent results were obtained using a $\sigma$ value of $0.2$ \AA.

\begin{figure}[tbh]
\centering
\includegraphics[width=0.8\textwidth]{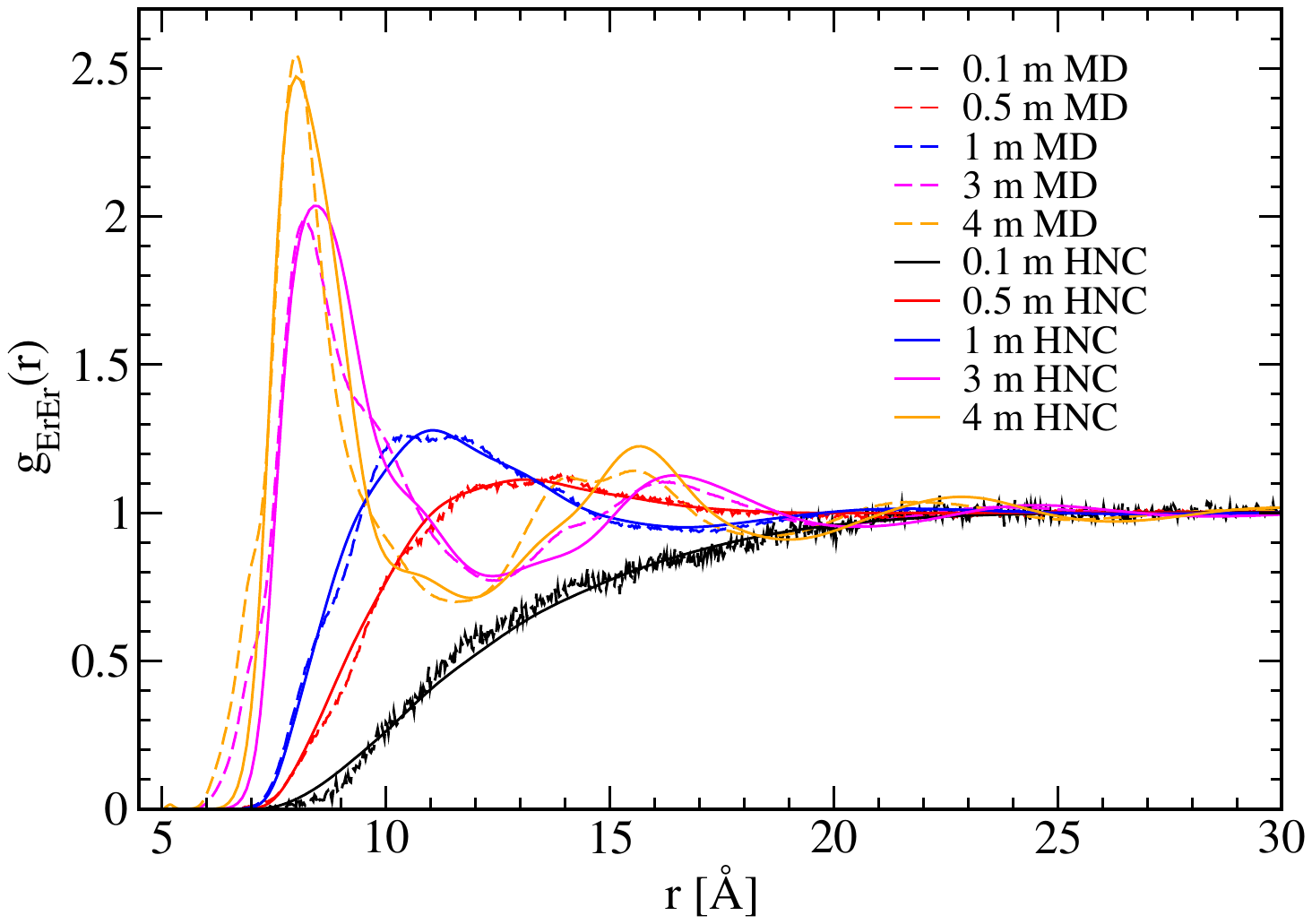}
\caption{Comparison of ErEr radial distribution functions from MD and HNC methods at different concentrations as described in the text. 
}
\label{compare-md-hnc}
\end{figure}

The assumptions we made, in the above procedures, were verified by directly comparing the results from  the OZ solver using the HNC closure (hereafter HNC results) and the MD results. For instance, Fig. \ref{compare-md-hnc} shows excellent agreements between MD and HNC results for the ErEr radial distribution functions at various concentrations.

\section{Definitions of static structure factors}

Consider a solution consisting of $n_i$ atoms for each species with a total number of $n$ atoms and an atomic number density of $\rho=n/V$.

\begin{equation}
S_{ij}(Q) = c_i \delta_{ij} + c_i c_j \rho \hat{h}_{ij}(Q)
\label{Sij}
\end{equation}
where $c_i=n_i/n$ is the atomic fraction of species $i$.

\begin{equation}
S_{NN}(Q) = \frac{1}{N} \langle \rho^N_{\mathbf{Q}} \rho^N_{\mathbf{-Q}} \rangle = \sum_i \sum_j S_{ij}(Q)
\label{SNN}
\end{equation}

\begin{equation}
S_{zz}(Q) = \frac{1}{N} \langle \rho^z_{\mathbf{Q}} \rho^z_{\mathbf{-Q}} \rangle = \sum_i \sum_j z_i z_j S_{ij}(Q)
\label{Szz}
\end{equation}
where $z_i$ is the charge for species $i$.

Making use of $S_{zz}(Q)$, one can define
\begin{equation}
\epsilon(Q) = \frac{1}{1-\frac{k^2_{\rm D} S_{zz}(Q)}{\zeta Q^2} }
\label{chi}
\end{equation}

\begin{equation}
\chi_\mathrm{L}(Q) = 1 - \frac{1}{\epsilon(Q)}
\label{chi}
\end{equation}

with $ k_{\rm D}^2 = 4 \pi \rho l_{\mathrm{B}} \zeta$ and $\zeta = \sum c_i z_i^2 $, $l_{\mathrm{B}} = e^2/(\epsilon k_{\mathrm{B}} T)$.

The measured X-ray intensity $I(Q)$ is related to the structure factor $S_{X}(Q)$
by
\begin{equation}
I(Q)\approx S_{X}(Q)-1=\rho \sum_{i,j}c_{i}c_{j}f_{ij}(Q)\hat{h}_{ij}(Q)
\end{equation}
where
\begin{equation}
f_{ij}(Q)=\frac{f_{i}(Q)f_{j}(Q)}{\left[{\displaystyle \sum_{i=1}^{N}}c_{i}f_{i}(Q)\right]^{2}}
\end{equation}
where $f_{i}(Q)$ is the X-ray form
factor for species $i$, and $N$ is the total number of atomic species in the system.

One may approximate the measured X-ray intensity by assuming that the form factors are constant and equal to the atomic number as 

\begin{equation}
S_{ZZ}(Q)= 
1 + \rho f \sum_{i,j} Z_{i}Z_{j}c_{i}c_{j}\hat{h}_{ij}(Q)
\label{SZZ_X}
\end{equation}
where
\begin{equation}
f=\frac{1}{\left[ \sum_i Z_i c_i  \right]^{2}}
\end{equation}
with $Z_i$ is the the atomic number for species $i$.

\section{HNC Root Structures for the ErCl$_3$ Electrolytes: $\chi(Q)$}

Following the discussions in the main text, we use the following functions to extract the roots structures from $\chi(Q)$:

\begin{equation}
 \chi_\mathrm{L}(Q) = \frac{\kappa_x^2}{Q^2 + \kappa_x^2 - Q^2 \tilde{\Sigma}(Q)}
 \label{gen_chi}
\end{equation}

For the auxiliary field model, we used 
\begin{equation}
\tilde{\Sigma}^\mathrm{Aux}(Q)= \frac{A^2}{1+g(Q)} 
 \label{aux_f}
\end{equation}
with $\displaystyle g(Q)=\sum_{i=1}^{4} b_i Q^{2i}$ and $\kappa_x = \kappa_s$, where the coefficients $A$, $b_i$ and $\kappa_s$ are determined during the analyses.

For the Pad$\acute{\mathrm{e}}$ function, we chose 
\begin{equation}
 \tilde{\Sigma}^\mathrm{Pad\acute{e}}(Q)= \frac{- (a + bQ^{2})}{1+cQ^{2}+dQ^{4}} 
 \label{pade_f}
\end{equation}
where $\kappa_x=\kappa_\mathrm{D}$ and $a$, $b$, $c$, and $d$ coefficients are determined during the analyses.

The root structures are extracted by first fitting $\chi(Q)$ in the low $Q$ region (up to about $2.5$ \AA$^{-1})$ to functions of type Eq. \ref{gen_chi} for each functional. The root structures are then obtained by finding the roots of the aforementioned functions making use of Muller's method in {\it mpmath} Python library, which is recommended for complex roots. In short, it starts with three initial assumptions of the root, and then constructing a parabola through these three points, and taking the intersection of the x-axis with the parabola to be the next approximation. This process continues until a root with the desired level of accuracy is found.
We use the initial guesses from the total correlation functions of $h_{++}(r)$ or $h_{+-}(r)$.

We also obtain the root structures for $\chi(Q)$ using the Teubner-Stey (TS) function presented below noting that the auxiliary field model and the Pad$\acute{\mathrm{e}}$ functions are generally able to capture the $\chi(Q)$ behavior in a wider range of $Q$-space.
\begin{equation}
 \chi^{\mathrm{TS}}_\mathrm{L}(Q)= \frac{4\pi A\left[\left(a_0^{2}-Q_0^{2}+Q^{2}\right)\cos[\delta]+2a_0 Q_o\sin[\delta]\right]}{\left[a_o^{2}+Q_o^{2}\right]^{2}+2\left[a_o^{2}-Q_o^{2}\right]Q^{2}+Q^{4}} + m Q + b
 \label{TS_f}
\end{equation}

\begin{figure}[tbh]
\centering
\includegraphics[width=0.8\textwidth]{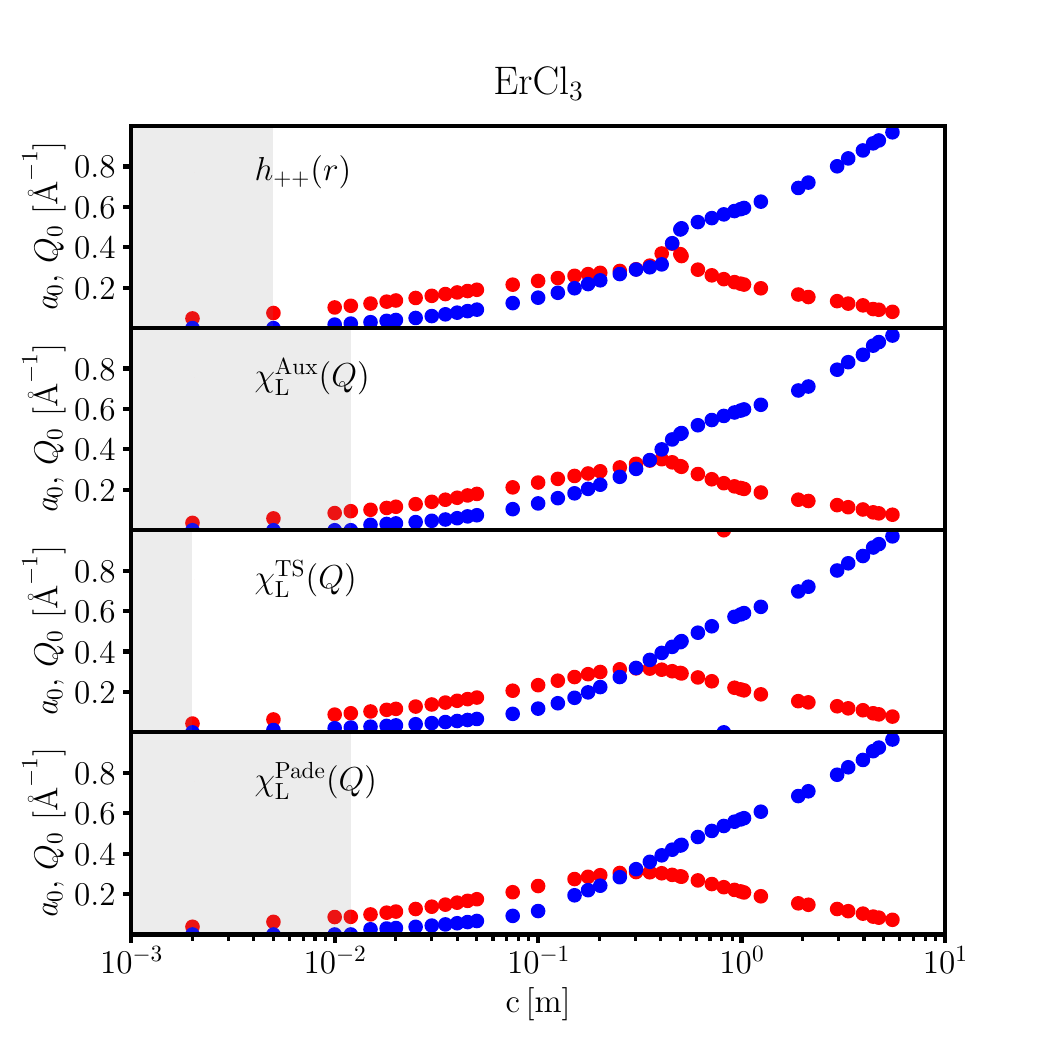}
\caption{The root structures for ErCl$_3$ using $\chi_\mathrm{L}(Q)$, where the roots are determined from the auxiliary functional (Eq. \ref{aux_f}), Pad\'e approximation (Eq. \ref{pade_f}), Teubner-Strey (TS) functional (Eq. \ref{TS_f}). The root structures are compared with the ones obtained from $h_{++}(r)$.  
}
\label{ercl3_root}
\end{figure}

\begin{figure}[tbh]
\centering
\includegraphics[width=0.8\textwidth]{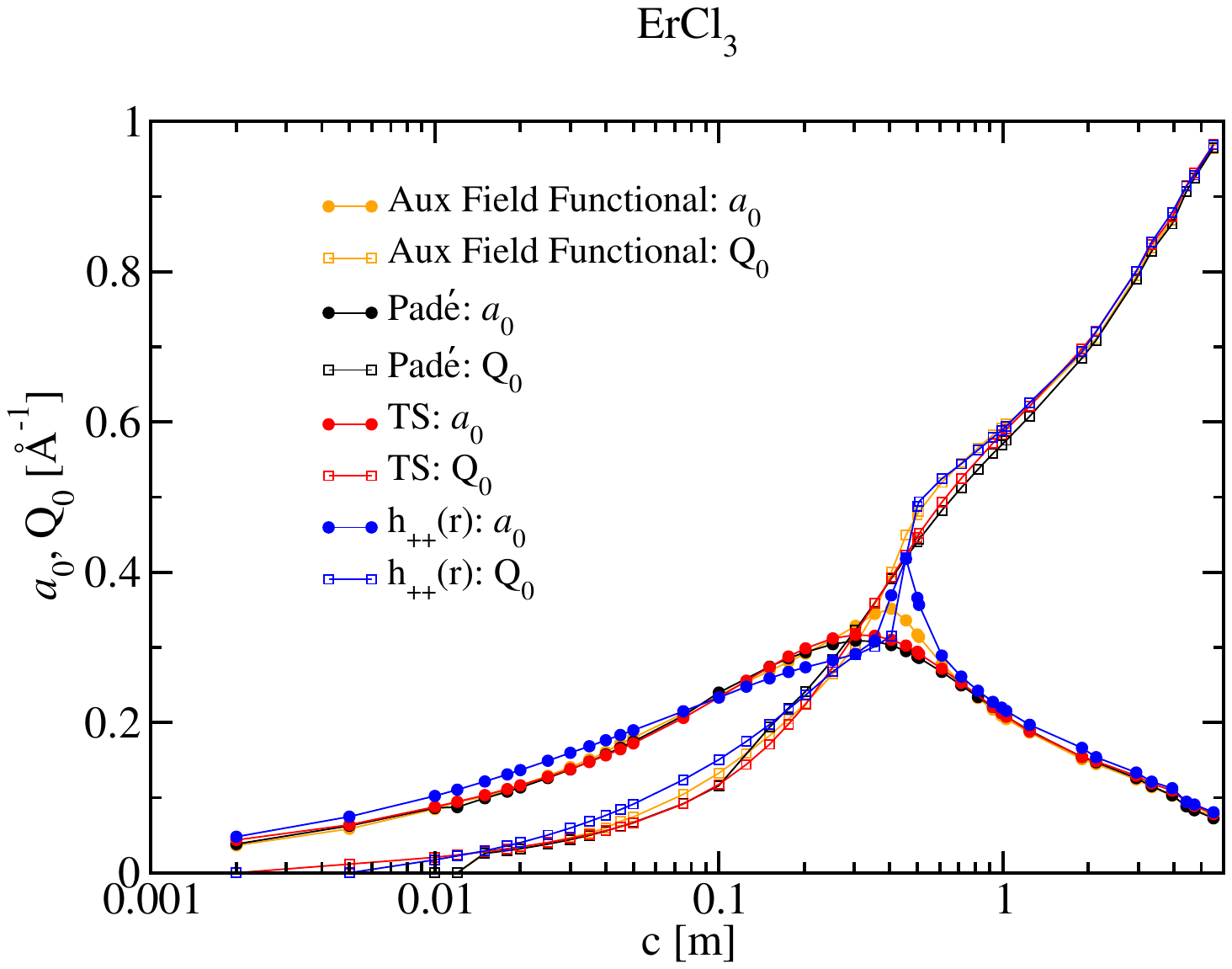}
\caption{The root structures for ErCl$_3$ using $\chi_\mathrm{L}(Q)$, where the roots are determined from the auxiliary functional (Eq. \ref{aux_f}), Pad\'e approximation (Eq. \ref{pade_f}), Teubner-Strey (TS) functional (Eq. \ref{TS_f}). The root structures are compared with the ones obtained from $h_{++}(r)$.  
}
\label{ercl3_root_busy}
\end{figure}

As can be seen in Figs. \ref{ercl3_root} and \ref{ercl3_root_busy}, for ErCl$_3$ all the root structures from the aforementioned functionals agree reasonably well. However, the auxiliary functional and Pad\'e functional can give the correct root structures over a larger region of $Q$-space than the ones from the TS functional.  

\clearpage
\section{Complex Root Structure for Various Hierarchies}

As discussed above and in the main text, we observe that the root structures for ErCl$_3$ significantly deviate from the traditional Kirkwood transition (KT). For SrCl$_2$ and CaCl$_2$, we observe slight deviations from the traditional KT while for NaCl the traditional picture almost holds. Below, we present the root structures from various hierarchies showing that they all show the root structure behaviors consistent with the traditional KT.  

\begin{figure}[tbh]
\centering
\includegraphics[width=0.8\textwidth]{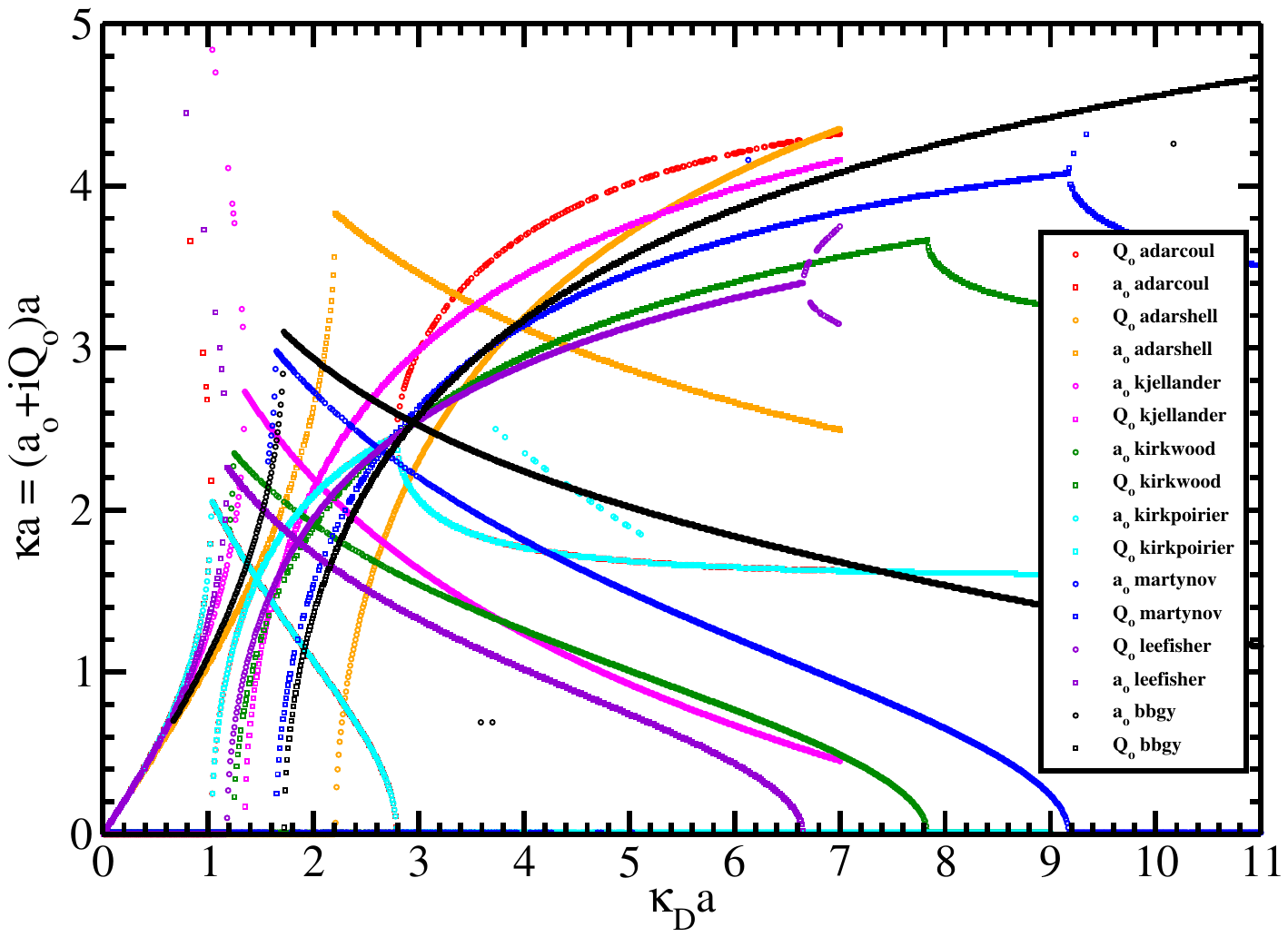}
\caption{A comparison of the complex root structure for several hierarchies showing similarities and differences. See Ref. \citen{Kirkwood1936,Kirkwood-1954,Singer1975} for kirkwood. See Ref. \citen{Kirkwood-1954} for kirkpiorier. See Ref. \citen{LeeFisher1997} for leefisher. See Refs. \citen{Kjellander1995,Kjellander2019} for kjellander. See Ref. \citen{Adar2019} for adarcoul and adarshell. See Refs. \citen{Outhwaite-1974,Singer1975} for bbgy and martynov.
}
\label{roots}
\end{figure}

\clearpage

\clearpage
\section{Related Fitting Procedures}

In the main text and the Supplementary Information, several signals are fitted to extract the $a_0$ and $Q_0$ values. In this Section, the fitting procedure is discussed in more details.

The X-ray signals, $I(Q)$, are fitted to TS distributions with linear background according to
\begin{equation}
I(Q) = 
\frac{4\pi A_0\left[\left(a_0^{2}-Q_o^{2}+Q^{2}\right)\cos[\delta]+2a_o Q_0\sin[\delta]\right]}{\left[a_o^{2}+Q_o^{2}\right]^{2}+2\left[a_o^{2}-Q_o^{2}\right]Q^{2}+Q^{4}} + m Q + b
\end{equation}
where $m$ and $n$ are the coefficients of the linear background and it is assumed that X-ray signal can be described by only one phase, $\delta$.

In $r$-space, the total correlation functions, $h(r)$, for various species are fitted to the damped oscillator functions of
\begin{equation}
h(r) = A_0 e^{-a_or}\cos(Q_o r-\delta)/r
\end{equation}

In HNC or MD calculations, one can get the response electrostatic potential as described in the main text. 
One may also extract the $a_0$ and $Q_0$ from the response electrostatic potentials according to
\begin{equation}
\phi_{j}^{\mathrm{resp}}(r)=-\frac{z_{j}e}{\varepsilon r}e^{-a_0r}
\left[\frac{\left(a_0^{2}-Q_0^{2}\right)\sin[Q_0r-\delta_{s}]+2a_0Q_0\cos[Q_0r-\delta_{s}]}{\left(a_0^{2}-Q_0^{2}\right)\sin[\delta_{s}]-2a_0Q_0\cos[\delta_{s}]}\right]
-\frac{z_{j}e}{\varepsilon r}
\label{eqn_resp}
\end{equation}
where $\delta_{s}=(\delta_{n}+\delta_{m})/2$ and $n$ and $m=$ 1 and 2 for cations and 2 and 3 for anions, respectively.

In addition, one can calculate the ionic atmosphere around a given ion $i$ in HNC or MD simulations via

\begin{equation}
S_0(r) = 4 \pi \int_0^r \rho_i(r') r'^2 dr' 
\label{atm_ion}
\end{equation}
where $\rho_j(r')$ is the charge density around the ion. 

Similarly, one may extract the $a_0$ and $Q_0$ values from $S_0(r)$ according to

\begin{equation}
S_0(r)=-z_{j}e[a\,+e^{-a_0r}\left(b\,\cos[Q_0r-\delta_s]+c\,\sin[Q_0r-\delta_s]\right)]
\label{eqn_S0}
\end{equation}
where
\begin{equation}
\begin{split}
&a=\frac{\left(a_0^{2}-Q_0^{2}\right)\sin[\delta_{s}]-2a_0Q_0\cos[\delta_{s}]}{\left(a_0^{2}-Q_0^{2}\right)\sin[\delta_{s}]+2a_0Q_0\cos[\delta_{s}]}\\
&b=\frac{2a_0Q_0+Q_0(a_0^{2}+Q_0^{2})r}{\left(a_0^{2}-Q_0^{2}\right)\sin[\delta_{s}]+2a_0Q_0\cos[\delta_{s}]}\\
&c=\frac{a_0^2-Q_0^2+a_0(a_0^{2}+Q_0^{2})r}{\left(a_0^{2}-Q_0^{2}\right)\sin[\delta_{s}]+2a_0Q_0\cos[\delta_{s}]}.
\end{split}
\end{equation}

\begin{figure*}[h]
\centering
\includegraphics[width=6.5in]{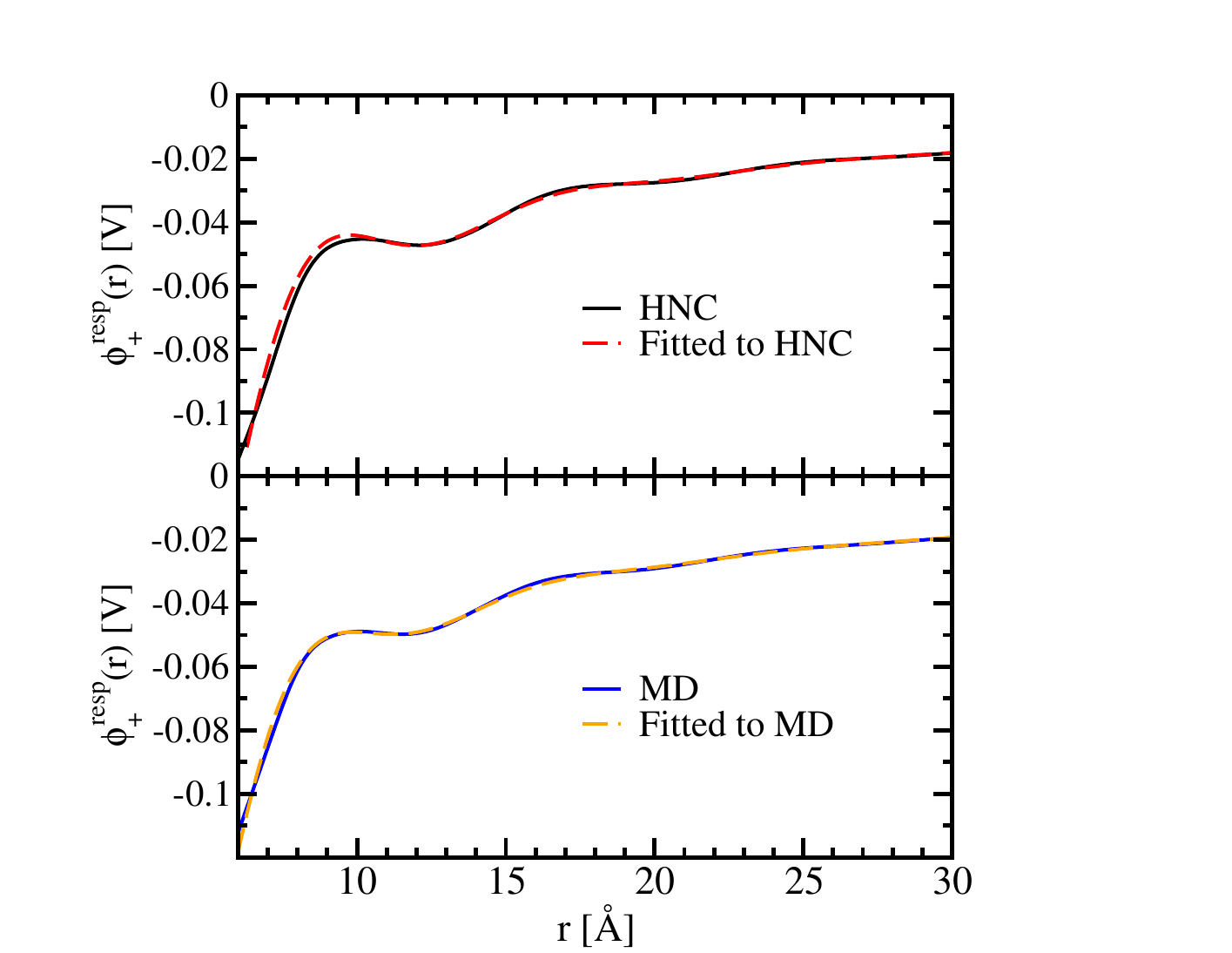}
\caption{Response electric potential $\phi_{+}^{\mathrm{resp}}(r)$ from MD and HNC calculations of ErCl$_3$ at 3 \emph{m} fitted with the analytic expression - see Eq. \ref{eqn_resp}.}
\label{fig:Sample_phi}
\end{figure*}

\begin{figure}[tbh]
\centering
\includegraphics[width=6.5in]{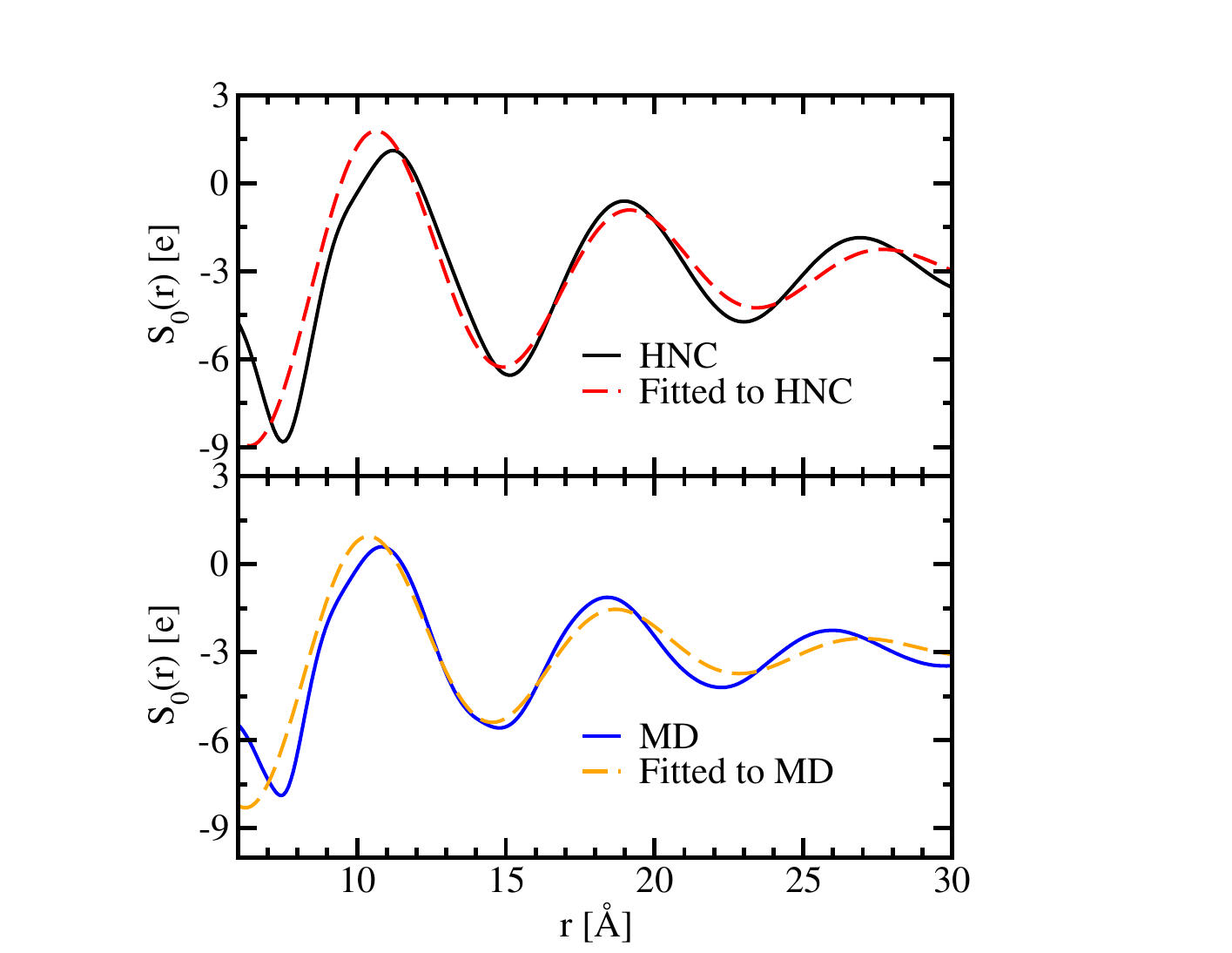}
\caption{Ion atmosphere $S_0(r)$ from MD and HNC calculations of ErCl$_3$ at 3 \emph{m} fitted with the analytic expression - see Eq. \ref{eqn_S0}.}
\label{fig:sample_KBI}
\end{figure}

Table \ref{fit_param1} presents the fitting parameters obtained from the experiment, MD, and HNC calculations using the above procedures for ErCl$_3$ at 3m. As can be seen, the $a_0$ and $Q_0$ values obtained from $I(Q)$ and $h_{++}(r)$ routes are in excellent agreement. 
Less agreements are observed when one uses $\phi_+^{\mathrm{resp}}(r)$ and $S_o(r)$ to extract $a_0$
 and $Q_0$ values; however, considering the uncertainties in the fitting procedures, they are still in good agreement. 
 Figures \ref{fig:Sample_phi} and \ref{fig:sample_KBI} show how well the fitting functions match the MD or HNC data.
 
\begin{table}[tbh]
\centering
\caption{Fitting parameters for the analytic functions provided above for $I(Q)$, $h_{++}(r)$, $\phi_+^{\mathrm{resp}}(r)$, and $S_o(r)$ from the experimental signals of ErCl$_3$ as well as MD and HNC calculations at $c=3\,m$. Linear background corrections were included in the estimations of $I(Q)$ and $\phi_+^{\mathrm{resp}}(r)$. To report $a_o$ and $Q_o$ in the main text, only $I(Q)$ and $h_{++}(r)$ functions were used.}
\vspace{0.3in}

\begin{tabular}{|c|c|cccccc|}
    \hline 
Method  &  Function & \multicolumn{6}{c|}{Fitting Parameters}\\
\hline 
&  $I(Q)$  & $a_o$[\AA$^{-1}$] & $Q_o$[\AA$^{-1}$] & $\delta$[rad] & $A$[\AA$^{-2}$] & $m$[\AA] & $b$  \\
     \hline
Expt &   & $0.162$ & $0.758$ & $0.95$ & $1.34$ & $14.20$ & $27.49$   \\
MD &   & $0.149$ & $0.805$ & $0.98$ & $-1.34$ & $9.40$ & $-46.58$   \\
\hline
&  $h_{++}(r)$  & $a_o$[\AA$^{-1}$] & $Q_o$[\AA$^{-1}$] & $\delta_{++}$[rad] & $A_0$[\AA] &  &  \\
     \hline
MD &   & $0.150$ & $0.824$ & $1.10$ & $17.13$ &  & \\
HNC  &   & $0.151$ & $0.782$ & $0.56$ & $24.48$ &  & \\
\hline
&  $\phi_+^{\mathrm{resp}}(r)$  & $a_o$[\AA$^{-1}$] & $Q_o$[\AA$^{-1}$] & $\delta_n$[rad] & $\delta_m$[rad] & $M$[V/\AA] &  $B$[V]  \\
     \hline
MD &    & $0.222$ & $0.767$ & $0.62$ & $2.09$ & $-1.0\times10^{-3}$ & $-8.5\times10^{-3}$ \\
HNC &    & $0.197$ & $0.757$ & $0.29$ & $2.84$ & $3.8\times10^{-6}$ & $-5.1\times10^{-5}$   \\
\hline 
&  $S_0(r)$   & $a_o$[\AA$^{-1}$] & $Q_o$[\AA$^{-1}$] & $\delta_s$[rad] &  & &  \\
\hline
MD &   & $0.190$ & $0.757$ & $1.58$ &  &  &  \\
HNC &   & $0.164$ & $0.738$ & $1.57$ &  &  & \\
     \hline
\end{tabular}
\label{fit_param1}
\end{table}

\clearpage
\section{Screening Lenghts and Periodicity}

\subsection{Ion-Ion vs. Water-Ion \& Water-Water Correlations}

\begin{figure}[tbh]
\centering
\includegraphics[width=1\textwidth]{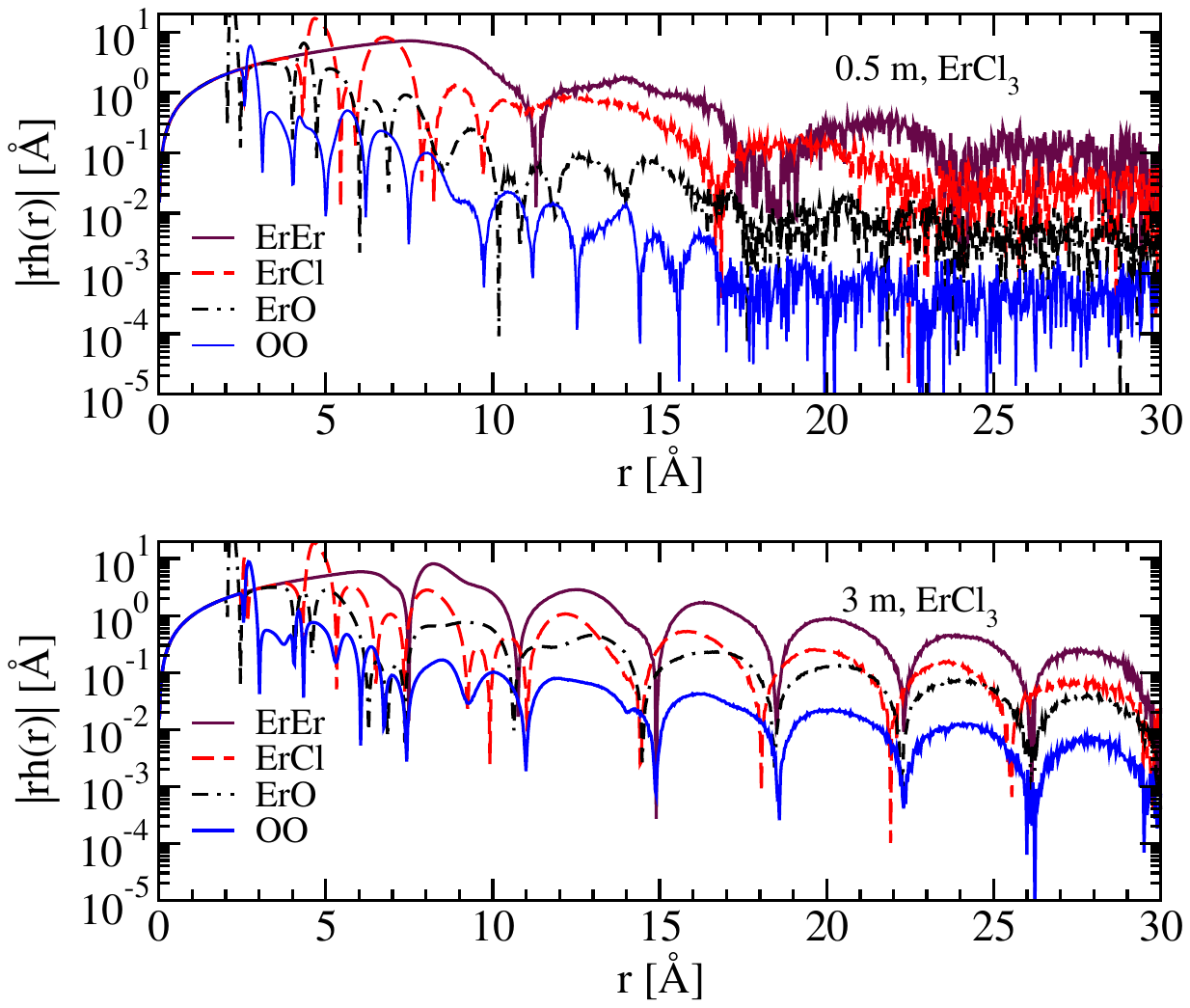}
\caption{Comparison of ion-ion correlations with ion-oxygen and oxygen-oxygen correlations for ErCl$_3$ at $0.5$ m and $3$ m from MD simulaitons.  
}
\label{ion-water}
\end{figure}

Figure \ref{ion-water} compares the ion-ion correlation behaviors with ion-water and water-water correlations in ErCl$_3$ electrolyte solutions at relatively low and high concentrations. The bottom panel of Fig. \ref{ion-water} shows that at a high ion concentration of $3$ m, the length scales associated with periodicity are almost consistent for ion-ion, ion-water, and water-water correlation. Nevertheless, the top panel of Fig. \ref{ion-water} at a lower ion concentration of $0.5$ m, such length scales for ion-ion correlations tend to be about twice as much as the ion-water and water-water correlations. 

\subsection{SrCl$_2$ \& CaCl$_2$}

\begin{figure*}[h]
\centering
\includegraphics[width=7.5in]{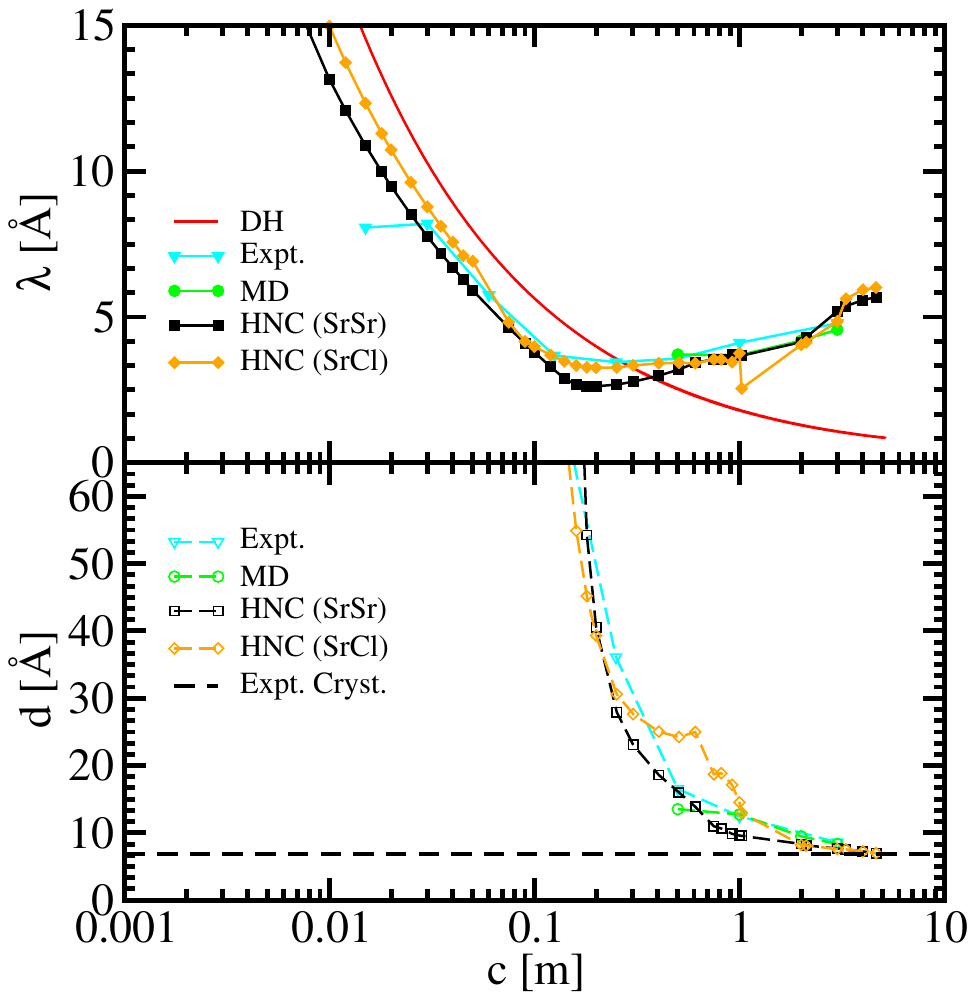}
\caption{A comparison of the $\lambda$ and $d$ from the experimental SAXS
spectra for aqueous SrCl$2$, MD simulation, HNC, and DH theory. The experimental Bragg spacing for the (100)-plane
crystalline SrCl$_2$ hexahydrate is $d=6.83$\,\AA.}
\label{fig:SrCl2_aoQo}
\end{figure*}

\begin{figure}[tbh]
\centering
\includegraphics[width=0.8\textwidth]{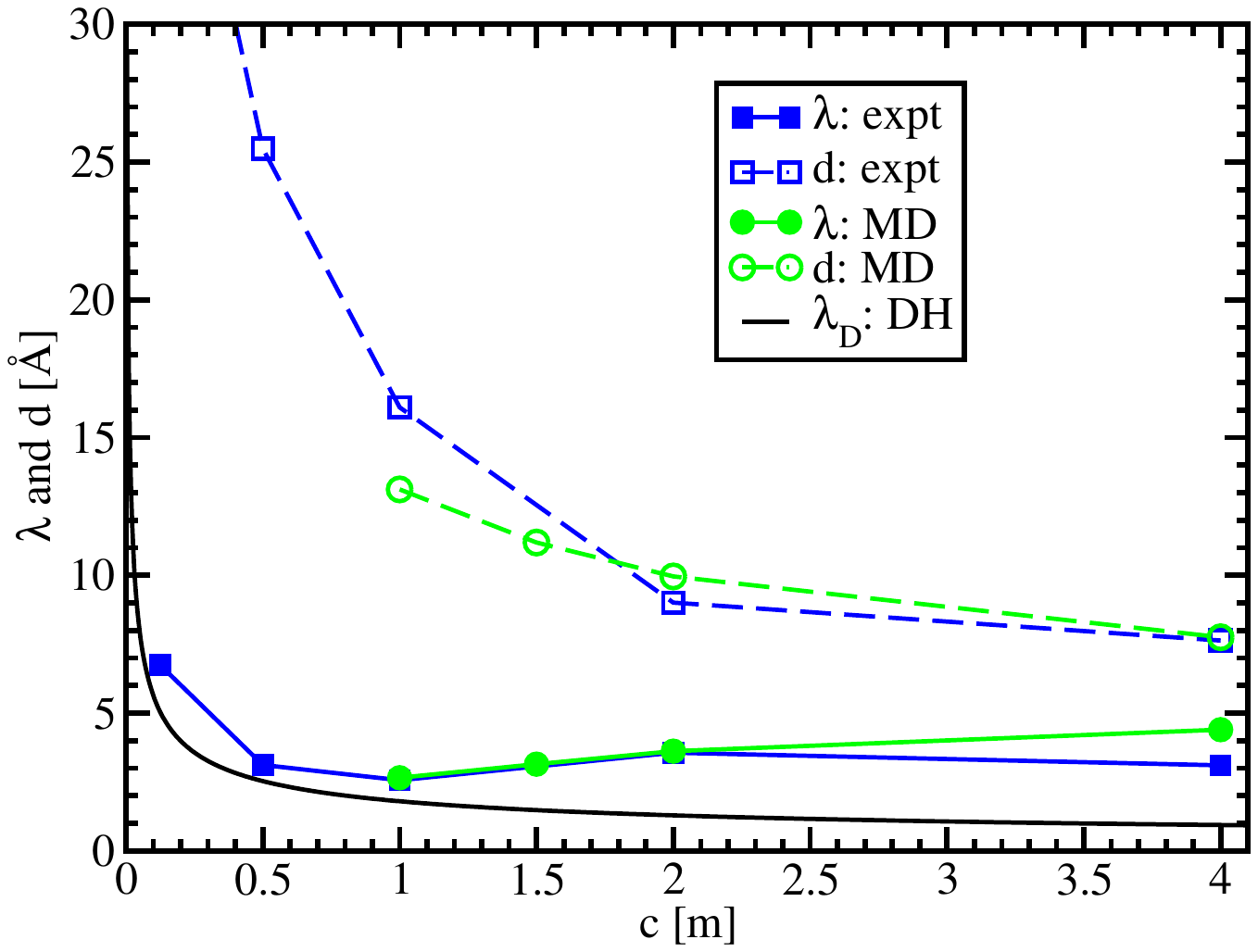}
\caption{A comparison of the $\lambda$ and $d$ from the experimental SAXS spectra for aqueous CaCl$_2$,
MD simulation, and DH theory. We note that the experimental Bragg spacing for the crystalline
CaCl$_2$ hexahydrate is $d = 6.825$  \AA.
}
\label{cacl2}
\end{figure}

Following the discussions in the main text, the $\lambda$ and $d$ values for SrCl$_2$ and CaCl$_2$ from various approaches are presented in Figs. \ref{fig:SrCl2_aoQo} and \ref{cacl2}.

\clearpage
\section{On the Convergence of MD simulations}

\begin{figure}[tbh]
\centering
\includegraphics[width=0.8\textwidth]{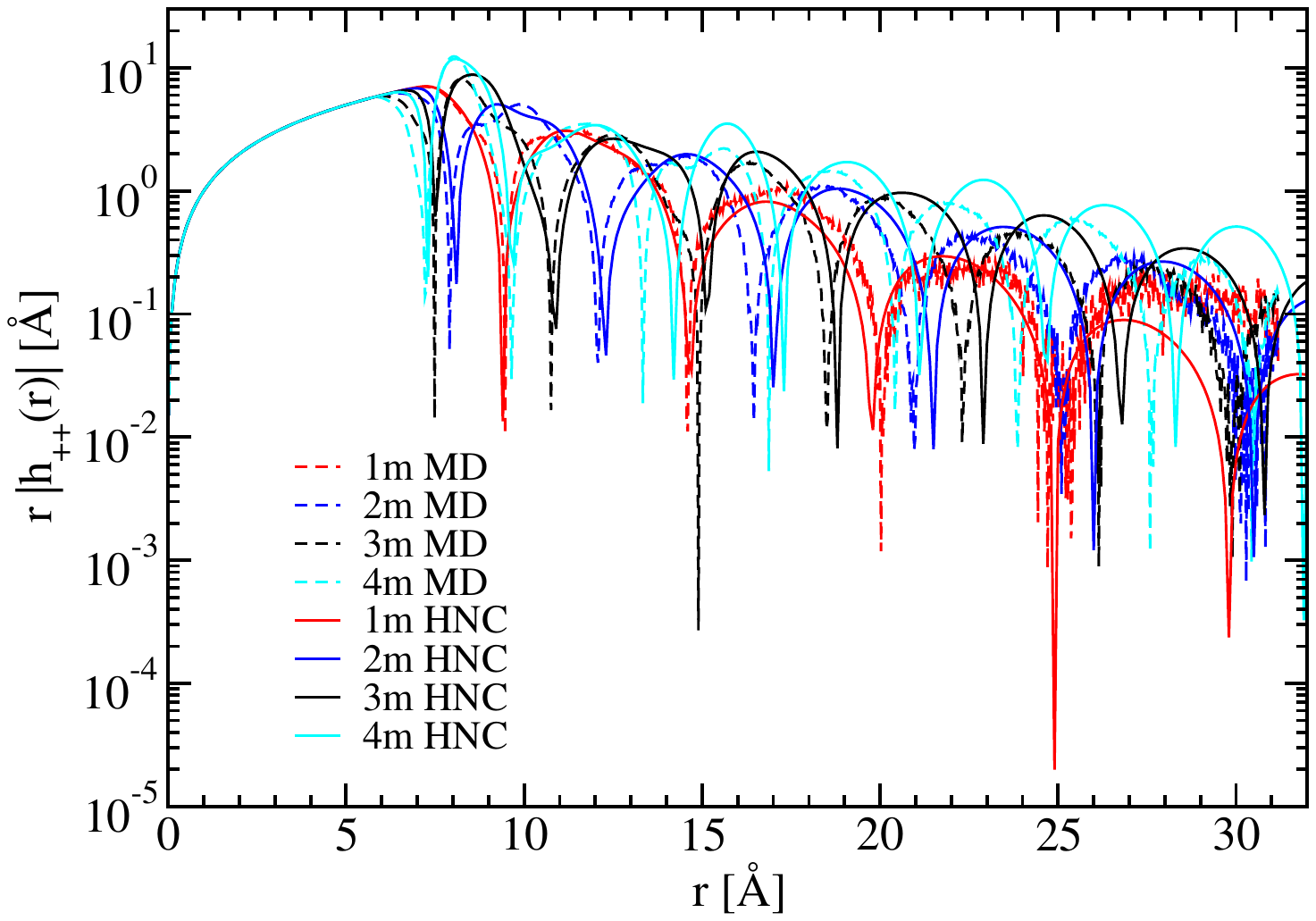}
\caption{Comparison of $r  |h(r)|$ for ErEr correlations from MD and HNC methods at various concentrations. 
}
\label{erer-ln_rhr_md-hnc}
\end{figure}

Figure \ref{erer-ln_rhr_md-hnc} shows the convergence of $r|h(r)|$ for ErEr from MD simulations at concentrations equal or higher than $1$ m. As can be seen, as the concentration increases, the noise in the tails of $r|h(r)|$ decrease.

\begin{figure}[tbh]
\centering
\includegraphics[width=0.8\textwidth]{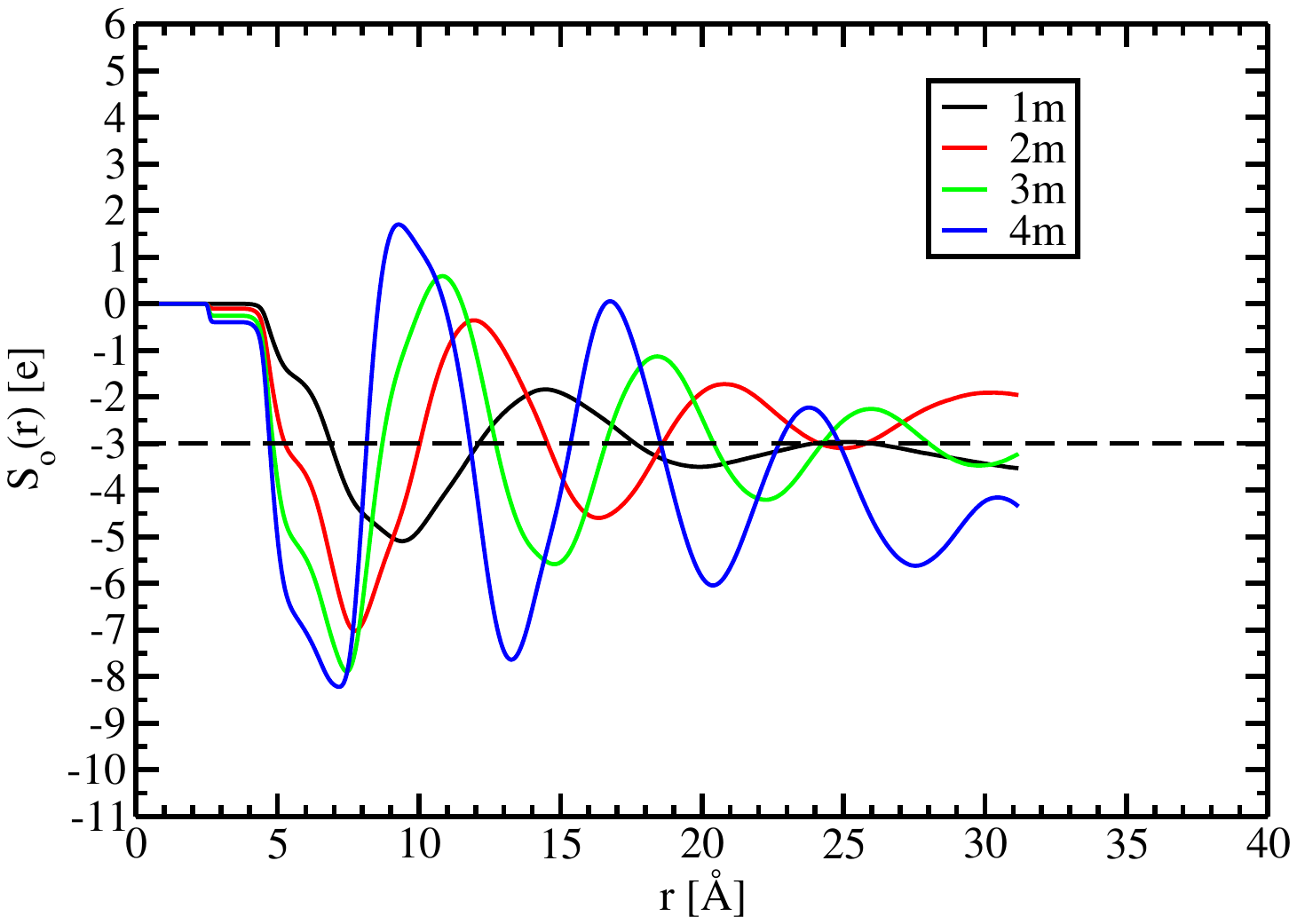}
\caption{Illustrating the convergence problems due to finite size effects in MD simulations to get $S_0(r)$ as defined in Eq. \ref{atm_ion}. Convergence of $S_0(r)$ is shown for various concentrations of ErCl$_3$.
}
\label{compare-md-kbi}
\end{figure}

Figure \ref{compare-md-kbi} shows the convergence of $S_0(r)$ (Eq. \ref{atm_ion}) from MD simulations for various concentrations of ErCl$_3$. As can be seen, the $S_0(r)$ at various distances oscillate around $-3$ e, but due to finite size effects in MD simulations there are deviations. 

\clearpage